\documentclass[a4paper,10pt,twoside]{cpc-hepnp}
\usepackage{multicol}
\usepackage{lastpage}
\usepackage{booktabs}
\usepackage{ragged2e}
\usepackage{graphicx}
\graphicspath{{figures/}}
\usepackage[utf8]{inputenc}
\usepackage[tbtags]{amsmath}
\usepackage{amsfonts}
\usepackage{amssymb}
\usepackage{multirow}
\usepackage{CJKutf8}
\usepackage{color}
\usepackage{acro}
\usepackage{adjustbox}
\usepackage{array}
\usepackage{natbib}
\usepackage{dblfnote}
\DFNalwaysdouble
\usepackage{slashed}
\usepackage{dcolumn}
\usepackage[colorlinks=true, linkcolor=red, citecolor=blue]{hyperref}
\def\({\left(}
\def\){\right)}
\def\[{\left[}
\def\]{\right]}
\def\be{\begin{eqnarray}}
	\def\ee{\end{eqnarray}}

\DeclareAcronym{GW}{
	short = GW ,
	long = gravitational wave ,
	short-plural = s %,  foreign-plural = {}
}
\DeclareAcronym{LIGO}{
	short = LIGO ,
	long = Laser Interferometer Gravitational-wave Observatory ,
	short-plural = s %,  foreign-plural = {}
}
\DeclareAcronym{LISA}{
	short = LISA ,
	long = Laser Interferometer Space Antenna ,
	short-plural =  %,  foreign-plural = {}
}
\DeclareAcronym{SKA}{
	short = SKA ,
	long = Square Kilometre Array ,
	short-plural =  %,  foreign-plural = {}
}
\DeclareAcronym{FLRW}{
	short = FLRW ,
	long = Friedmann-Lemaître-Robertson-Walker ,
	short-plural =  %,  foreign-plural = {}
}
\DeclareAcronym{TT}{
	short = TT ,
	long = transverse and traceless ,
	short-plural =  %,  foreign-plural = {}
}

\begin{document}
%\fancyhead[c]{\small Chinese Physics C~~~Vol. 37, No. 1 (2013) 010201}
%\fancyfoot[C]{\small010201-\thepage}

%\footnotetext[0]{Received 14 March 2009}

	\title{Note on gauge invariance of second order cosmological perturbations}

\author{%
	      Zhe Chang,$^{1,2}$
	\email{changz@ihep.ac.cn}
		 Sai Wang,$^{1,2}$
	\email{wangsai@ihep.ac.cn (Corresponding author)}
	     and Qing-Hua Zhu$^{1,2}$ 
	\email{zhuqh@ihep.ac.cn (Corresponding author)}}
\maketitle

\address{$^1$Theoretical Physics Division, Institute of High Energy Physics, Chinese Academy of Sciences, Beijing 100049, People's Republic of China\\
	$^2$School of Physical Sciences, University of Chinese Academy of Sciences, Beijing 100049, People's Republic of China\\
}

\begin{abstract}
We study the gauge invariant cosmological perturbations up to second order. We show that there are infinite families of gauge invariant variables at both of the first and second orders. The conversion formulae among different families are shown to be described by a finite number of bases that are gauge invariant. For the second order cosmological perturbations induced by the first order scalar perturbations, we explicitly represent the equations of motion of them in terms of the gauge invariant Newtonian, synchronous and hybrid variables, respectively. 
\end{abstract}

%\begin{keyword}
%	Cosmological perturbation theory, physics of the early universe, gauge invariance
%\end{keyword}

%\footnotetext[0]{\hspace*{-3mm}\raisebox{0.3ex}{$\scriptstyle\copyright$}2013
%	Chinese Physical Society and the Institute of High Energy Physics
%	of the Chinese Academy of Sciences and the Institute
%	of Modern Physics of the Chinese Academy of Sciences and IOP Publishing Ltd}%

\begin{multicols}{2}
\section{Introduction}

The gauge invariance is known as an important concept in the cosmological perturbation theory. 
At first order, the gauge invariant variables were introduced for the first time by Bardeen in 1980s {\cite{Bardeen:1980kt}}. Since then, different gauge-invariant formalisms were studied  \cite{PhysRevD.40.1804,Bruni:1992dg,Sopuerta:2003rg,Bruni:1996im,Sonego:1997np,Nakamura:2006rk}.
They have been widely utilized to study the cosmological fluctuations \cite{Mukhanov:1990me}, for example, the anisotropies in the cosmic microwave background and the large scale structures, which have been precisely measured in the past decades \cite{Zyla:2020zbs}. 
For the second order cosmological perturbations, the gauge invariance has also been studied recently, and a quantity of gauge invariant variables have been constructed \cite{Bruni:1996im,Sonego:1997np,Matarrese:1997ay,Malik:2008im,Nakamura:2006rk,Domenech:2017ems,Gong:2019mui,Lyth:2005du,Bruni:2011ta,Bruni:1999et,Yoo:2017svj}. 
In addition, it is believed that the gauge invariance is available to study the higher order cosmological perturbations \cite{Nakamura:2014kza}. 

However, the things changed recently when one studied the second order cosmological perturbations, including but not limited to the second order gravitational waves \cite{Mollerach:2003nq,Ananda:2006af,Baumann:2007zm,Assadullahi:2009jc,Saga:2014jca}, which are induced by the first order scalar perturbations \cite{Alabidi:2012ex,Alabidi:2013lya,Inomata:2018epa,Orlofsky:2016vbd,Ben-Dayan:2019gll,Kohri:2018awv,Espinosa:2018eve,Wang:2019kaf,Kapadia:2020pir,Kapadia:2020pnr,Cai:2018dig,Yuan:2019wwo,Yuan:2020iwf,PhysRevD.81.023517,Bugaev:2010bb,Saito:2008jc,Saito:2009jt,Nakama:2016enz,Cai:2019jah}. 
The energy density spectrum of the second order gravitational waves as a physical observable have recently been involved in gauge issue \cite{Inomata:2019yww,Yuan:2019fwv,Giovannini:2020qta,Hwang:2017oxa,Lu:2020diy,DeLuca:2019ufz,Tomikawa:2019tvi}. 
One possible explanation is related to the gauge fixing, which may give rise to unknown fictitious perturbations \cite{Giovannini:2020qta}. 
Therefore, one may resolve this problem by constructing the gauge invariant variables \cite{Bruni:1996im,Sonego:1997np,Matarrese:1997ay,Malik:2008im,Nakamura:2006rk,Domenech:2017ems,Gong:2019mui,PhysRevD.40.1804,Bruni:1992dg,Sopuerta:2003rg}. 
The other explanation is related to the definition of the physical observable, which has been suggested to be defined in the synchronous frame \cite{DeLuca:2019ufz}. 
However, in such a framework, the concept of gauge invariance was suggested to be necessarily abandoned, since it is impossible to truly construct the gauge invariant second order synchronous variables \cite{Bertschinger:1993xt}. 

However, we would show that the gauge invariance could be preserved when one studies the second order cosmological perturbations, in particular, the second order gravitational waves. 
On the one hand,  the power spectrum as the physical observable should be gauge invariant. 
Otherwise, it could take an arbitrary value if we choose an appropriate gauge fixing, e.g., the synchronous gauge \cite{Lifshitz:1945du}, as reviewed in Ref.~\cite{Bertschinger:1993xt}. 
On the other hand, the gauge invariant synchronous variables can be reasonably defined for the induced gravitational waves at second order, though they are ill-defined for the second order scalar and vector perturbations. 
In this sense, the energy density spectrum of the second order gravitational waves could be well defined. % as will be studied in a companion paper \cite{Chang:2020pre}. 

 In this work, we will investigate the gauge invariance of the second order cosmological perturbations by following the Lie derivative method \cite{Matarrese:1997ay,Sonego:1997np,Nakamura:2006rk,Malik:2008im,Bruni:1996im}. 
The gauge invariant variables of any order can be systematically constructed in such a method. 
We will study the gauge invariant Newtonian (i.e., Bardeen's) and synchronous variables at first and second orders. 
In particular, for the first time, we will construct the so-called gauge invariant hybrid variables, i.e., Newtonian at first order while synchronous at second order, and vice versa. 
Further, we will find the conversion formulae among different families of gauge invariant variables. 
Finally, we will derive the gauge invariant equations of motion for the second order cosmological perturbations by adopting the scalar-vector-tensor decomposition \cite{Nakamura:2011xy}. 
To accomplish it, we will decompose the gauge invariant perturbed Einstein field equations into the scalar, vector and tensor components. 

The remainder of the paper is arranged as follows. In Section \ref{II}, we introduce the gauge
transformations and the gauge invariant variables following the Lie derivative method. In Section \ref{III}, we revisit the gauge invariant first order variables in such a method. In Section \ref{IV}, we present explicit expressions of the gauge invariant second
order variables and study the conversion formulae among different families of gauge invariant variables. 
In Section \ref{V}, the equations of motion of the cosmological perturbations are presented in terms of the gauge invariant Newtonian, synchronous and hybrid variables. Finally, the conclusions and
discussions are summarized in Section \ref{VI}.
%In addition, we list a total of six appendices. 

\section{Gauge transformation and Gauge invariant variables}\label{II} 

\subsection{Gauge transformation of an arbitrary tensor} 

A gauge transformation for a perturbed quantity in space-time starts from an
infinitesimal transformation of coordinate $x^{\mu} \rightarrow
\tilde{x}^{\mu}$. The expansion of $\tilde{x}^{\mu}$ up to
second order is given by
\begin{equation}
  \tilde{x}^{\mu} = x^{\mu} + \xi^{(1) \mu} + \frac{1}{2} \xi^{(2) \mu}
  +\mathcal{O} (\xi^{(3)}) \label{1},
\end{equation}
where $\xi^{(1) \mu}$ and $\xi^{(2) \mu}$ are the first and second order
expansions of $\tilde{x}^{\mu}$, respectively. 
For a generic tensor $\mathcal{Q}$, to second order, the infinitesimal transformations are shown to be \cite{Bruni:1996im} 
\begin{subequations}
\begin{eqnarray}
%  \tilde{\mathcal{Q}}^{(0)} & = & \mathcal{Q}^{(0)},  \label{6}\\
  \tilde{\mathcal{Q}}^{(1)} & = & \mathcal{Q}^{(1)} +\mathcal{L}_{\xi_1}
  \mathcal{Q}^{(0)},  \label{7}\\
  \tilde{\mathcal{Q}}^{(2)} & = & \mathcal{Q}^{(2)} + 2\mathcal{L}_{\xi_1}
  \mathcal{Q}^{(1)} + (\mathcal{L}_{\xi_2} +\mathcal{L}_{\xi_1}^2)
  \mathcal{Q}^{(0)},  \label{8}
\end{eqnarray}\label{A1}
\end{subequations}
where we have let $\xi_1^{\mu} \equiv \xi^{(1) \mu}$ and $\xi_2^{\mu} \equiv
\xi^{(2) \mu} - \xi^{(1)\nu} \partial_{\nu} \xi^{(1) \mu}$, 
$\mathcal{Q}^{(n)}$ denotes the $n$-th order perturbation of the tensor
$\mathcal{Q}$, and $\mathcal{L}_{\xi}$ is Lie derivative along
the infinitesimal vector $\xi$. The degree of freedom of the transformation is the same for the first and the second order perturbations. The $\xi_1$ determines the $\tilde{\mathcal{Q}}^{(1)}$. Once the $\xi_1$ is fixed, thus the $\xi_2$ determines the $\tilde{\mathcal{Q}}^{(2)}$. These formulae are based on that the $\xi_1$ is set to be the same order of tensor perturbation $\mathcal{Q}^{(1)}$. It could simplify the formalism of perturbation theory. For a general infinitesimal transformation, there is not relevance between $\xi_1$ and $\mathcal{Q}^{(1)}$ on orders.
In Appendix \ref{A}, an introduction to the Lie derivative is shown. 
We briefly summarize the derivations of Eqs.~(\ref{A1}) in Appendix \ref{A-1}. 
%As shown in Eq.~(\ref{6}), the gauge transformation we discussed is limited to the infinitesimal transformation that has no influence on the background. 

\subsection{Gauge invariant metric perturbations}

Based on Eqs.~(\ref{7}) and (\ref{8}), the infinitesimal transformations of
the first and second order metric perturbations are presented as
\begin{subequations}
\begin{eqnarray}
  \tilde{g}_{\mu \nu}^{(1)} & = & g_{\mu \nu}^{(1)} +\mathcal{L}_{\xi_1}
  g_{\mu \nu}^{(0)},  \label{9}\\
  \tilde{g}^{(2)}_{\mu \nu} & = & g_{\mu \nu}^{(2)} + 2\mathcal{L}_{\xi_1}
  g_{\mu \nu}^{(1)} + (\mathcal{L}_{\xi_2} +\mathcal{L}_{\xi_1}^2) g_{\mu
  \nu}^{(0)} .  \label{10}
\end{eqnarray}
\end{subequations}
If we introduce the gauge invariant metric perturbations via $g_{\mu
\nu}^{(\mathrm{GI}, i)} \equiv g_{\mu \nu}^{(i)} - \mathbb{C}_{\mu \nu}^{(i)}$ ($i = 1,
2$ for example), by making use of
Eqs.~(\ref{9}) and (\ref{10}), we could rewrite the counter terms $\mathbb{C}_{\mu \nu}^{(i)}$ in
terms of the infinitesimal vectors $X^{\mu}$ and $Y^{\mu}$, namely, 
\begin{subequations}
\begin{eqnarray}
  g_{\mu \nu}^{(\mathrm{\rm{GI}, 1)}} & = & g_{\mu \nu}^{(1)} -\mathcal{L}_X
  g_{\mu \nu}^{(0)},  \label{11}\\
  g^{(\mathrm{\rm{GI}, 2)}}_{\mu \nu} & = & g_{\mu \nu}^{(2)} -
  2\mathcal{L}_X g_{\mu \nu}^{(1)} - (\mathcal{L}_Y -\mathcal{L}_X^2) g_{\mu
  \nu}^{(0)},  \label{12}
\end{eqnarray}
\end{subequations}
where
\begin{subequations}
\begin{eqnarray}
  \tilde{X}^{\mu}  & = & X^{\mu}+ \xi_1^{\mu},  \label{13}\\
  \tilde{Y}^{\mu}  & = & Y^{\mu}+ \xi_2^{\mu} + [\xi_1, X]^{\mu} .  \label{14}
\end{eqnarray}
\end{subequations}
We present the derivations of Eqs.~(\ref{11})--(\ref{14}) in Appendix \ref{B}. 
These formulae could also be found in
Refs.~{\cite{Nakamura:2006rk,Nakamura:2004rm,Nakamura:2019zbe}}.
Recently, they have been used in cosmology
{\cite{Nakamura:2006rk,Nakamura:2019zbe}} and
in the post-Newtonian formalism {\cite{Hohmann:2019qgo}}.

Based on Eqs.~(\ref{13}) and (\ref{14}), the infinitesimal vectors $X^{\mu}$ and $Y^{\mu}$ could be independent of the metric perturbations in principle. At first order, Bardeen has constructed the gauge invariant variables in terms of the metric perturbations {\cite{Bardeen:1980kt}}. Therefore, we limit our investigations to the case that both $X^\mu$ and $Y^\mu$ are expressed in terms of the metric perturbations  in the following. 

\subsection{Gauge invariant perturbations of the energy-momentum tensor}

In the aspect of matter perturbations, we adopt the infinitesimal
transformation in Eqs.~(\ref{7}) and (\ref{8}) to obtain the perturbed energy-momentum
tensors, i.e., 
\begin{subequations}
\begin{eqnarray}
  \tilde{T}_{\mu \nu}^{(1)} & = & T_{\mu \nu}^{(1)} +\mathcal{L}_{\xi_1}
  T_{\mu \nu}^{(0)},  \label{9-1}\\
  \tilde{T}^{(2)}_{\mu \nu} & = & T_{\mu \nu}^{(2)} + 2\mathcal{L}_{\xi_1}
  T_{\mu \nu}^{(1)} + (\mathcal{L}_{\xi_2} +\mathcal{L}_{\xi_1}^2) T_{\mu
  \nu}^{(0)} .  \label{10-1}
\end{eqnarray}
\end{subequations}
The gauge invariant perturbed energy-momentum tensors can be given by
\begin{subequations}
\begin{eqnarray}
  T_{\mu \nu}^{(\mathrm{\rm{GI}, 1)}} & = & T_{\mu \nu}^{(1)} -\mathcal{L}_X
  T_{\mu \nu}^{(0)},  \label{15}\\
  T^{(\mathrm{\rm{GI}, 2)}}_{\mu \nu} & = & T_{\mu \nu}^{(2)} -
  2\mathcal{L}_X T_{\mu \nu}^{(1)} - (\mathcal{L}_Y -\mathcal{L}_X^2) T_{\mu
  \nu}^{(0)},  \label{15-1}
\end{eqnarray}
\end{subequations}
%Once the variables $X^{\mu}$ and $Y^{\mu}$ are known from the metric perturbations [Eq.~(\ref{11}) and (\ref{12})], the $T_{\mu \nu}^{(\mathrm{GI},\alpha)}$ can be obtained, directly.
In the above formulae, it should be noticed that the energy-momentum tensors are
not in a special status. 
In general, any of the gauge invariant tensors could be
formulated in the same form as Eqs.~(\ref{15}) and (\ref{15-1}), i.e.,
\begin{subequations}
\begin{eqnarray}
  \mathcal{Q}^{(\mathrm{\rm{GI}, 1)}} & = & \mathcal{Q}^{(1)} -\mathcal{L}_X
  \mathcal{Q}^{(0)},  \label{15-2}\\
  \mathcal{Q}^{(\mathrm{\rm{GI}, 2)}} & = & \mathcal{Q}^{(2)} -
  2\mathcal{L}_X \mathcal{Q}^{(1)} - (\mathcal{L}_Y -\mathcal{L}_X^2)
  \mathcal{Q}^{(0)} . \label{15-3}
\end{eqnarray}
\end{subequations}

\subsection{Gauge invariant Einstein field equations} \label{IIC}

%As shown by Mukhanov {\cite{mukhanov_theory_1992}} and Nakamura
%{\cite{nakamura_second order_2020}}, w
We have already known that the
gauge invariant perturbed Einstein field equations have the same form
as the conventional ones {\cite{Mukhanov:1990me,Nakamura:2019zbe}}. 
This can be explicitly shown by expanding the Einstein field equations, $G_{\mu \nu} = \kappa T_{\mu \nu}$, namely, 
\begin{eqnarray}
  0 & = & G_{\mu \nu} - \kappa T_{\mu \nu} \nonumber\\
  & \approx & G^{(0)}_{\mu \nu} - \kappa T_{\mu \nu}^{(0)} + G^{(1)}_{\mu
  \nu} - \kappa T_{\mu \nu}^{(1)} + \frac{1}{2}  (G_{\mu \nu}^{(2)} - \kappa
  T_{\mu \nu}^{(2)}) \nonumber\\
  & = & G_{\mu \nu}^{(0)} - \kappa T_{\mu \nu}^{(0)} \nonumber\\
  &  & +
  (G^{(\mathrm{\rm{GI}, 1)}} - \kappa T_{\mu \nu}^{(\mathrm{\rm{GI}, 1)}}
  +\mathcal{L}_X (G_{\mu \nu}^{(0)} - \kappa T_{\mu \nu}^{(0)})) \nonumber\\
  &  & + \frac{1}{2}  (G_{\mu \nu}^{(\mathrm{\rm{GI}, 2)}} - \kappa T_{\mu
  \nu}^{(\mathrm{\rm{GI}, 2)}} + 2\mathcal{L}_X (G_{\mu \nu}^{(1)} - \kappa
  T_{\mu \nu}^{(1)}) \nonumber\\
  &  & + (\mathcal{L}_Y -\mathcal{L}_X^2) (G_{\mu \nu}^{(0)} -
  \kappa T^{(0)}_{\mu \nu})),  \label{15a}
\end{eqnarray}
where $G_{\mu \nu}$ is Einstein tensor, $\kappa \equiv 8 \pi G$, and $G$ is the
gravitational constant. 
We could separate Eq.~(\ref{15a}) order by order to obtain
\begin{subequations}
\begin{eqnarray}
  G_{\mu \nu}^{(0)} & = & \kappa T_{\mu \nu}^{(0)},  \label{16}\\
  G_{\mu \nu}^{(\mathrm{\rm{GI}, 1)}} & = & \kappa T_{\mu
  \nu}^{(\mathrm{\rm{GI}, 1)}},  \label{17}\\
  G_{\mu \nu}^{(\mathrm{\rm{GI}, 2)}} & = & \kappa T_{\mu
  \nu}^{(\mathrm{\rm{GI}, 2)}} . \label{20c}
\end{eqnarray}
\end{subequations}
%It shows that the gauge invariant perturbed Einstein field equations are established at each order.
The gauge invariant perturbed Einstein
tensor $G_{\mu \nu}^{(\mathrm{\rm{GI}, n)}}$ takes the same form as the conventional one $G_{\mu \nu}^{(n)}$. It can be obtained by substituting the metric
perturbations $g_{\mu \nu}^{(n)}$ with the gauge invariant ones
$g_{\mu \nu}^{(\mathrm{\rm{GI}, n)}}$. 
The same situation is also true for the perturbed energy-momentum tensor $T_{\mu\nu}^{(\mathrm{\rm{GI}, n})}$. 

In fact, the above conclusion is obvious. It can be understood as follows.  
As known, the perturbed Einstein tensor $G_{\mu \nu}^{(n)}$ are defined with the metric perturbations $g_{\mu\nu}^{(n)}$. 
Upon the gauge transformation, the transformed perturbed Einstein tensor $\tilde{G}^{(n)}_{\mu \nu}$ should be written in terms of the transformed metric perturbations $\tilde{g}_{\mu \nu}^{(n)}$. 
Based on this, since we notice that $g_{\mu
\nu}^{(\mathrm{\rm{GI}, n)}}$ in Eqs.~(\ref{11}) and (\ref{12}) formally take
 the same form as the gauge transformations, i.e., $\tilde{g}_{\mu \nu}^{(n)}$ in Eqs.~(\ref{9}) and (\ref{10}), the gauge invariant perturbed Einstein tensor $G_{\mu \nu}^{(\mathrm{\rm{GI}, n)}}$ should be defined with the gauge invariant metric perturbations $g_{\mu\nu}^{(\mathrm{\rm{GI}, n)}}$.
%As an example, we will explicitly show the above conclusion at first order in Appendix~\ref{B-1}. 

\section{Gauge invariant variables for the first order cosmological metric perturbations}\label{III}

The gauge invariant first order variables were first proposed by Bardeen
{\cite{Bardeen:1980kt}}.
% and widely used by Mukhanov
%{\cite{mukhanov_theory_1992}}. 
As was known, Bardeen's gauge invariant
variables have the same form as the metric perturbations in
Newtonian gauge. In this section, we will reproduce Bardeen's formulae and
further show that there are  infinite families of gauge invariant variables that
are allowed.

\subsection{Gauge transformations of the first order metric perturbations}
In flat \ac{FLRW} space-time, the metric takes the form of 
\begin{equation}
  g_{\mu \nu}^{(0)} \mathrm{d} x^{\mu} \mathrm{d} x^{\nu} = a^2 (\eta)  (-
  \mathrm{d} \eta^2 + \delta_{ij} \mathrm{d} x^i \mathrm{d} x^j),
\end{equation}
where $\eta$ and $a(\eta)$ are the conformal time and the scale factor of the Universe, respectively. The spatial curvature of the space-time is zero.
The metric perturbations of $n$-th order can take the form of
\end{multicols}
\hspace{0mm}\vspace{5mm}\ruleup
\begin{equation}
  g_{\mu \nu}^{(n)} = a^2 \left( \begin{array}{ccc}
    - 2 \phi^{(n)} &  & \partial_i b^{(n)} + \nu_i^{(n)}\\
    \partial_i b^{(n)} + \nu_i^{(n)} &  & - 2 \psi^{(n)}
    \delta_{ij} + 2 \partial_i \partial_j {e}^{(n)} + \partial_i
    c_j^{(n)} + \partial_j c_i^{(n)} + h_{ij}^{(n)}
  \end{array} \right) \label{18},
\end{equation}
\vspace{5mm}\ruledown
\begin{multicols}{2}\hspace{-6.5mm}
where $\phi^{(n)}$, $\psi^{(n)}$, $b^{(n)}$, ${e}^{(n)}$ are
scalar perturbations, $\nu_i^{(n)}$ and $c_j^{(n)}$ are vector
perturbations and $h_{ij}^{(n)}$ are tensor perturbations. 
For tensor and vector perturbations, the transverse or traceless conditions should be satisfied as follows
\begin{subequations}
\begin{eqnarray}
    \partial_i \nu^{i (n)} & = & 0,\\
    \partial_i c^{i (n)} & = & 0,\\
    \delta^{jk} \partial_k h_{ij}^{(n)} & = & 0,\\
    \delta^{ij} h_{ij}^{(n)} & = & 0.
\end{eqnarray}
\end{subequations}
These variables can be introduced via scalar-vector-tensor
decomposition, which is summarized in Appendix \ref{C}.

Following Eq.~(\ref{9}), the gauge transformations of the scalar, vector and tensor perturbations are explicitly shown as 
\begin{subequations}
\begin{eqnarray}
\tilde{g}^{(1)}_{00}   &=& - 2 a^2  \tilde{\phi}^{(1)}\nonumber\\
 &=& - 2 a^2 \phi^{(1)} - 2 a^2  (  \partial_0 + \frac{\dot{a}}{a} ) \xi_1^0,  \label{20}\\
\tilde{g}^{(1)}_{0i}  & = & a^2  (\partial_i  \tilde{b}^{(1)} + \tilde{\nu}_i^{(1)})\nonumber\\
& = & a^2  (\partial_i b^{(1)} + \nu_i^{(1)}) + a^2  (\delta_{ij} \partial_0 \xi_1^j -
  \partial_i \xi^0_1), \\
\tilde{g}^{(1)}_{ij}   &=  & a^2  (- 2 \tilde{\psi}^{(1)} \delta_{ij} + 2 \partial_i \partial_j 
  \tilde{e}^{(1)} + \partial_i  \tilde{c}_j^{(1)} + \partial_j 
  \tilde{c}_i^{(1)} + \tilde{h}_{ij}^{(1)})  \nonumber \\
  &=&   a^2  (- 2 \psi^{(1)}
  \delta_{ij} + 2 \partial_i \partial_j {e}^{(1)} + \partial_i c_j^{(1)} +
  \partial_j c_i^{(1)} + h_{ij}^{(1)})  \nonumber\\
  &  & + a^2  ( (\delta_{ik}
  \partial_j + \delta_{jk} \partial_i) (\xi^k_{1, T} + \delta^{kl} \partial_l
  \xi_{1,S}) \nonumber\\
  &  &  + \frac{2 \dot{a}}{a} \delta_{ij} \xi^0_1 ) .  \label{22}
\end{eqnarray}
\end{subequations}
Here, the spatial part of $\xi_1^{\mu}$ has been decomposed as $\xi^i_1 =
\xi^i_{1, T} + \delta^{ij} \partial_j \xi_{1,S}$, where we have the transverse part
$\xi_{1, T}^i$ and the longitudinal part $\xi_{1,S}$. Using Eqs.~(\ref{20})--(\ref{22})
and the scalar-vector-tensor decomposition, we rewrite the gauge transformations of
the variables of metric perturbations as follows
\begin{subequations}
\begin{eqnarray}
    \tilde{\phi}^{(1)} & = & \phi^{(1)} + \left( \partial_0 +
    \frac{\dot{a}}{a} \right) \xi_1^0,\\
    \tilde{b}^{(1)} & = & b^{(1)} + \partial_0 \xi_{1,S} - \xi_1^0,\\
    \tilde{\nu}_i^{(1)} & = & \nu_i^{(1)} + \delta_{ij} \partial_0 \xi^j_{1,
    T},\\
    \tilde{\psi}^{(1)} & = & \psi^{(1)} - \frac{\dot{a}}{a} \xi_1^0,\\
    \tilde{e}^{(1)} & = & {e}^{(1)} + \xi_{1,S},\\
    \tilde{c}_i^{(1)} & = & c_i^{(1)} + \delta_{ik} \xi^k_{1, T},\\
    \tilde{h}_{ij}^{(1)} & = & h_{ij}^{(1)} .
\end{eqnarray} \label{23}
\end{subequations}
%The scalar, vector and tensor perturbations are shown to be not entangled with other parts of the metric perturbations upon the gauge transformations.

In the following, we will introduce the gauge invariant variable of metric perturbations by
making use of Eq.~(\ref{23}). Since the gauge invariant metric perturbations could take the same form as  Newtonian (or synchronous) gauge, we call them the gauge invariant Newtonian (or synchronous) metric perturbations for the sake of presentations.

\subsection{Gauge invariant first order Newtonian variables}\label{III.1}

%Based on Eqs.~(\ref{9}) and (\ref{11}), we can find that the gauge invariant
%metric perturbation $g_{\mu \nu}^{(\mathrm{\rm{GI}, 1)}}$ takes a similar
%form as the formula of $\tilde{g}_{\mu \nu}^{\mathrm{(1)}}$ in the gauge
%transformations. Thus, f
Based on Eq.~(\ref{11}), we could obtain the gauge invariant Newtonian metric perturbations, i.e., 
\begin{subequations}
\begin{eqnarray}
  g_{00}^{(\mathrm{\rm{GI}, 1)}} & = & - 2 a^2 \Phi^{(1)},  \label{24-0}\\
  g_{0 i}^{(\mathrm{\rm{GI}, 1)}} & = & a^2 V_i^{(1)},  \label{24}\\
  g_{ij}^{(\mathrm{\rm{GI}, 1)}} & = & a^2  (- 2 \Psi^{(1)} \delta_{ij} +
  H_{ij}^{(1)}),  \label{24-1}
\end{eqnarray}
\end{subequations}
where the gauge invariant variables are defined as
\begin{subequations}
\begin{eqnarray}
    \Phi^{(1)} & = & \phi^{(1)} - ( \partial_0 + \frac{\dot{a}}{a}
    ) X^0,\\
    \Psi^{(1)} & = & \psi^{(1)} + \frac{\dot{a}}{a} X^0,\\
    V^{(1)}_i & = & \nu_i^{(1)} - \delta_{ij} \partial_0 X^j_T,\\
    H_{ij}^{(1)} & = & h_{ij}^{(1)},
\end{eqnarray}\label{25}
\end{subequations}
and we have decomposed $X^i = : X_T^i + \delta^{ij} \partial_j X_S$, i.e., into the
transverse part and the longitudinal one.  As expected that $X^{\mu}$
is expressed in terms of the metric perturbations $g^{(1)}_{\mu \nu}$, we
can derive its formula from Eq.~(\ref{24}),
\begin{eqnarray}
   a^2 V_i^{(1)} & = & g_{0 i}^{(1)}
  -\mathcal{L}_X g_{0 i}^{(0)} \nonumber\\
  & = & a^2  (\partial_i (b^{(1)} - \partial_0 X_S + X^0) \nonumber\\
  & &+ (\nu_i^{(1)} -
  \delta_{ji} \partial_0 X^j_T)) . 
\end{eqnarray}
Since the vector perturbation is transverse, it leads to
\begin{equation}
  b^{(1)} - \partial_0 X_S + X^0 = 0. \label{27}
\end{equation}
By making use of Eq.~(\ref{24-1}), namely, 
\begin{eqnarray}
  && a^2  (- 2 \Psi^{(1)} \delta_{ij} +
  H_{ij}^{(1)}) = g_{ij}^{(1)} -\mathcal{L}_X g_{ij}^{(0)} \nonumber\\
  & &= a^2  ( - 2 \delta_{ij}  ( \psi^{(1)} + \frac{\dot{a}}{a}
  X^0  ) + 2 \partial_i \partial_j ({e}^{(1)} - X_S) \nonumber\\
  & &\hspace{3mm} + (\delta^k_j \partial_i
  + \delta^k_i \partial_j) (c_k^{(1)} - \delta_{lk} X^l_T) + h_{ij}^{(1)}), 
\end{eqnarray}
and the scalar-vector-tensor decomposition, we obtain 
\begin{subequations}
\begin{eqnarray}
  {e}^{(1)} - X_S & = & 0,  \label{29}\\
  c_k^{(1)} - \delta_{lk} X_T^l & = & 0.  \label{30}
\end{eqnarray}
\end{subequations}
%This is due to  and the transverse and traceless conditions. 
Based on Eqs.~(\ref{27}), (\ref{29}) and (\ref{30}),
$X^{\mu}$ can be given by
\begin{subequations}
\begin{eqnarray}
    X^0 & = & \partial_0 {e}^{(1)} - b^{(1)},\\
    X^i & = & \delta^{ik}  (c_k^{(1)} + \partial_k {e}^{(1)}) .
\end{eqnarray}\label{31}
\end{subequations}
Therefore, we rewrite the gauge invariant variables as
\begin{subequations}
\begin{eqnarray}
  \Phi^{(1)} & = & \phi^{(1)} - \frac{1}{a} \partial_0  (a (\partial_0 {e}^{(1)}
  - b^{(1)})),  \label{31-1}\\
  \Psi^{(1)} & = & \psi^{(1)} - \frac{\dot{a}}{a}  (\partial_0 {e}^{(1)} -
  b^{(1)}), \\
  V_i^{(1)} & = & \nu_i^{(1)} - \partial_0 c_i^{(1)},  \label{31-3}\\
  H_{ij}^{(1)} & = & h_{ij}^{(1)} .  \label{31-2}
\end{eqnarray}
\end{subequations}
This implies that the gauge invariant variables proposed by Bardeen
{\cite{Bardeen:1980kt}} can be reproduced by making use of Lie
derivative method \cite{Nakamura:2019zbe}. 

\subsection{Gauge invariant first order synchronous variables}

Based on Eq.~(\ref{11}),  the gauge invariant synchronous metric perturbations take the form of
\begin{subequations}
\begin{eqnarray}
    g_{00}^{(\mathrm{\rm{GI}, 1)}} & = & 0, \label{36-0}\\
    g_{0 i}^{(\mathrm{\rm{GI}, 1)}} & = & 0,\\
    g_{ij}^{(\mathrm{\rm{GI}, 1)}} & = & a^2  (- 2 \Psi^{(1)} \delta_{ij} +
    2 \partial_i \partial_j E^{(1)} \nonumber\\
    & &+ \partial_i C_j^{(1)} +
    \partial_j C_i^{(1)} + H_{ij}^{(1)}), \label{36}
\end{eqnarray}
\end{subequations}
where the gauge invariant variables are defined as
\begin{subequations}
\begin{eqnarray}
    \Psi^{(1)} & = & \psi^{(1)} + \frac{\dot{a}}{a} X^0,\\
    E^{(1)} & = & {e}^{(1)} - X_S,\\
    C_i^{(1)} & = & c_i^{(1)} - \delta_{ik} X_T^k,\\
    H_{ij}^{(1)} & = & h_{ij}^{(1)} .
\end{eqnarray}\label{37}
\end{subequations}
In this case, $X^{\mu}$ can be determined by the form $g_{0
\mu}^{(\mathrm{\rm{GI}, 1)}} = 0$ and the scalar-vector-tensor decomposition.
The result is given by
\begin{subequations}
\begin{eqnarray}
    X^0 &=&  \frac{1}{a}  \int \mathrm{d} \eta \{a \phi^{(1)} \},\\
    X^j & = & \delta^{ji}  \int \mathrm{d} \eta \{ \nu_i^{(1)} +
    \partial_i b^{(1)} + \frac{1}{a}  \int \mathrm{d} \eta' \{a \partial_i
    \phi^{(1)} \} \} . 
\end{eqnarray}\label{39}
\end{subequations}
Then we rewrite the gauge invariant variables in the form of
\begin{subequations}
\begin{eqnarray}
  \Psi^{(1)} & = & \psi^{(1)} + \frac{\dot{a}}{a^2}  \int \mathrm{d} \eta \{a
  \phi^{(1)} \},  \label{44}\\
  E^{(1)} & = & {e}^{(1)} + \int \mathrm{d} \eta \{ b^{(1)} +
  \frac{1}{a}  \int \mathrm{d} \eta' \{a \phi^{(1)} \} \}, \\
  C_i^{(1)} & = & c_i^{(1)} - \int \nu_i^{(1)} \mathrm{d} \eta, \\
  H_{ij}^{(1)} & = & h_{ij}^{(1)} .  \label{47}
\end{eqnarray}
\end{subequations}
%As the gauge invariant $g_{\mu \nu}^{(\mathrm{\rm{GI}, 1)}}$
%[Eq.~(\ref{36})] takes the same form as synchronous gauge, we might call it
%synchronous gauge invariant metric perturbations for the sake of presentation.
The above approach has been used to study the scalar cosmological perturbations in
Ref.~{\cite{Uggla:2011hs}}. As suggested in Ref.~\cite{DeLuca:2019ufz}, the gauge invariant synchronous variables are not unique due to the  indefinite integral in Eq.~(\ref{39}). In this sense, it might be difficult to define observables with the gauge invariant synchronous variables in Eqs.~(\ref{44})--(\ref{47}).

\subsection{Conversion among different families of gauge invariant first order variables}

In general, the gauge invariant variables are not limited to the forms in
Eqs.~(\ref{24-0})--(\ref{24-1}) and (\ref{36-0})--(\ref{36}). There are other families of gauge invariant variables (see reviews in Refs.~\cite{Malik:1998ai,Malik:2008im}). For two different families of
the gauge invariant first order variables of metric perturbations $g_{\mu
\nu}^{(\mathrm{\rm{GI}, A, 1)}}$ and $g_{\mu \nu}^{(\mathrm{\rm{GI}, B,
1)}}$, the conversion between them can be derived from
\begin{eqnarray}
  g_{\mu \nu}^{(\mathrm{\rm{GI}, A, 1)}} - g_{\mu \nu}^{(\mathrm{\rm{GI},
  B, 1)}} & = & (g_{\mu \nu}^{(1)} -\mathcal{L}_{X^A} g_{\mu \nu}^{(0)}) -
  (g_{\mu \nu}^{(1)} -\mathcal{L}_{X^B} g_{\mu \nu}^{(0)}) \nonumber\\
  & = & \mathcal{L}_{(X^B - X^A)} g_{\mu \nu}^{(0)} .  \label{59-1}
\end{eqnarray}
If we let $Z_1^{\mathrm{AB}} \equiv X^A - X^B$, Eq.~(\ref{59-1}) can be
rewritten as
\begin{equation}
  g_{\mu \nu}^{(\mathrm{\rm{GI}, A, 1)}} = g_{\mu \nu}^{(\mathrm{\rm{GI},
  B, 1)}} -\mathcal{L}_{Z_1^{\mathrm{AB}}} g_{\mu \nu}^{(0)} . \label{59a}
\end{equation}
The variable $Z_1^{\mathrm{AB}}$ was mentioned by Nakamura
\cite{Nakamura:2019zbe}. It relates two different families of
gauge invariant first order variables of metric perturbations. 
In this work, we further show that the
infinitesimal vector $Z_1$ can be expressed as a linear combination of
the gauge invariant variables $\mathcal{A}^{(1)}$, $\mathcal{B}^{(1)}$ and
$\mathcal{C}^{(1)}_i$, namely,
\begin{equation}
  Z^{\mu, \mathrm{AB}}_1 = \hat{N}_1^{\mu} \mathcal{A}^{(1)} + \hat{N}_2^{\mu}
  \mathcal{B}^{(1)} + \hat{N}_3^{\mu i} \mathcal{C}_i^{(1)}, \label{60}
\end{equation}
where $\hat{N}^{\mu}_{1}$, $\hat{N}^{\mu}_{2}$ and $\hat{N}^{\mu j}_{3}$ are arbitrary
linear operators that are irrelative to any perturbations, and we have the gauge invariant variables,
\begin{subequations}
\begin{eqnarray}
  \mathcal{A}^{(1)} & \equiv & \partial_0  \big( \frac{a^2}{\dot{a}}
  \psi^{(1)} \big) + a \phi^{(1)}, \\
  \mathcal{B}^{(1)} & \equiv & \partial_0 {e}^{(1)} - b^{(1)} +
  \frac{a}{\dot{a}} \psi^{(1)}, \\
  \mathcal{C}_i^{(1)} & \equiv & \nu_i^{(1)} - \partial_0 c_i^{(1)} . 
\end{eqnarray}
\end{subequations}
The above expressions of $\mathcal{A}^{(1)}$, $\mathcal{B}^{(1)}$ and
$\mathcal{C}^{(1)}_i$ can be obtained by making use of Eq.~(\ref{23}).

The existence of  infinite families of gauge invariant variables can also be
indicated by the infinite number of choices of $X^{\mu}$. To be specific, we can extend the expression
of Eq.~(\ref{31}) to be
\begin{subequations}
\begin{eqnarray}
  X^0 & = & \partial_0 {e}^{(1)} - b^{(1)} + Z^0_1,  \label{50}\\
  X^j & = & \delta^{jk}  (c_k^{(1)} + \partial_k {e}^{(1)}) + Z^j_1 . 
  \label{51}
\end{eqnarray}
\end{subequations}
%One can check that $X^{\mu}$ [Eqs.~(\ref{50}) and (\ref{51})] still
%satisfies the constraint condition [Eq.~(\ref{13})], since $\mathcal{A}$,
%$\mathcal{B}$ and $\mathcal{C}_i$ vanish when evaluating $\tilde{X} - X$. 
In this way, the gauge invariant metric perturbations could take a general form of
\begin{subequations}
\begin{eqnarray}
    g_{00}^{(\mathrm{\rm{GI}, 1)}} & = & - 2 a^2 \Phi^{(1)},\\
    g_{0 i}^{(\mathrm{\rm{GI}, 1)}} & = & a^2  (\partial_i B^{(1)}
    + V_i^{(1)}),\\
    g_{ij}^{(\mathrm{\rm{GI}, 1)}} & = & a^2  (- 2 \Psi^{(1)} \delta_{ij} +
    2 \partial_i \partial_j E^{(1)} \nonumber\\
    & & + \partial_i C_j^{(1)} +
    \partial_j C_i^{(1)} + H_{ij}^{(1)}),
\end{eqnarray}\label{52}
\end{subequations}
where the gauge invariant variables are defined as
\begin{subequations}
\begin{eqnarray}
    \Phi^{(1)} & = & \phi^{(1)} - \frac{1}{a} \partial_0  (a (\partial_0
    {e}^{(1)} - b^{(1)} + Z^0_1)),\\
    B^{(1)} & = & Z^0_1 - \partial_0 \Delta^{- 1} \partial_j Z^j_1\\
    \Psi^{(1)} & = & \psi^{(1)} + \frac{\dot{a}}{a}  (\partial_0 {e}^{(1)} -
    b^{(1)} + Z^0),\\
    E^{(1)} & = & \Delta^{- 1} \partial_j Z^j_1,\\
    V_i^{(1)} & = & \nu_i^{(1)} - \partial_0 c_i^{(1)} - (\delta_{ik} -
    \partial_i \Delta^{- 1} \partial_k) \partial_0 Z^k_1,\\
    C_i^{(1)} & = & - (\delta_{ik} - \partial_i \Delta^{- 1}
    \partial_k) Z^k_1,\\
    H_{ij}^{(1)} & = & h_{ij},
\end{eqnarray}
\end{subequations}
and $\Delta^{- 1}$ is the inverse Laplacian operator on the background. 
%One might find that the
%gauge invariant metric perturbations [Eqs.~(\ref{52})] take the same form as
%the metric perturbations without gauge choices [Eq.~(\ref{18})].

As an example, we consider that $g_{\mu
\nu}^{(\mathrm{\rm{GI}, A, 1)}}$ and $g_{\mu \nu}^{(\mathrm{\rm{GI}, B,
1)}}$ are synchronous in Eqs.~(\ref{44})--(\ref{47}) and Newtonian in Eqs.~(\ref{31-1})--(\ref{31-2}) variables, respectively.
In this case, we obtain the explicit expression of $Z_1^{\mu, \mathrm{AB}}$ as follows
\begin{subequations}
\begin{eqnarray}
  Z^{0, \mathrm{AB}}_1 & = & \frac{1}{a}  \int \mathcal{A}^{(1)} \mathrm{d}
  \eta -\mathcal{B}^{(1)},  \label{59b}\\
  Z^{j, \mathrm{AB}}_1 & = & \delta^{jk} \partial_k  \int \mathrm{d} \eta
  \{ \frac{1}{a}  \int \mathcal{A}^{(1)} \mathrm{d} \eta' \} \nonumber\\
  & &-
  \delta^{jk} \partial_k  \int \mathcal{B}^{(1)} \mathrm{d} \eta + \delta^{ji}
  \int \mathcal{C}_i^{(1)} \mathrm{d} \eta .  \label{60b}
\end{eqnarray}
\end{subequations}
Therefore, these two families of gauge invariant variables of metric perturbations are related via the expressions of the linear operators
$\hat{N}_1^{\mu}$, $\hat{N}_2^{\mu}$ and $\hat{N}_3^{\mu i}$. 
%However, it should be noticed that two different  of gauge invariant metric perturbations are not related by gauge transformations.

\section{Gauge invariant variables for the second order cosmological metric perturbations}\label{IV}

%Recent investigations about scalar induced gravitational waves have possibly
%deduced a gauge dependence of the energy density spectrum
%{\cite{hwang_gauge_2017,gong_analytic_2019,tomikawa_gauge_2020,yuan_scalar_2020,lu_gauge_2020,inomata_gauge_2020}}.
%It also leads to the discussions about which gauges are the observables
%defined in {\cite{luca_gauge_2020}}. Therefore, it should be important to consider
%whether the second order cosmological perturbations are compatible with a
%gauge-invariance scenario.

%The gauge invariant second order variables are nothing new and were reviewed in Ref.~\cite{malik_cosmological_2009}. 
In this section, the gauge invariant second order variables in cosmology will be
derived in the framework of Lie derivative method and scalar-vector-tensor decomposition. Previous similar studies can be found in Ref.~\cite{Malik:2008im,Nakamura:2019zbe}.  Further, we will show the conversion formulae among different families of gauge invariant variables. For illustrations, we will consider the 
gauge invariant metric perturbations that are Newtonian and synchronous, respectively. 

\subsection{Gauge transformations of the second order metric perturbations}

Based on Eq.~(\ref{10}),
the gauge transformation of the second order metric perturbations is presented explicitly as follows
\end{multicols}
\hspace{0mm}\vspace{5mm}\ruleup
\begin{subequations}
\begin{eqnarray}
  \tilde{g}^{(2)}_{00} & = & - 2 a^2  \tilde{\phi}^{(2)} \nonumber\\
  & = & - 2 a^2 \phi^{(2)} + 2\mathcal{L}_{\xi_1} g_{00}^{(1)} +
  (\mathcal{L}_{\xi_2} +\mathcal{L}_{\xi_1}^2) g_{00}^{(0)},  \label{55}\\
  \tilde{g}_{0 i}^{(2)} & = & a^2  (\partial_i  \tilde{B}^{(2)} +
  \tilde{\nu}_i^{(2)}) \nonumber\\
  & = & g_{00}^{(2)} + 2\mathcal{L}_{\xi_1} g_{0 i}^{(1)} +
  (\mathcal{L}_{\xi_2} +\mathcal{L}_{\xi_1}^2) g_{0 i}^{(0)} \nonumber\\
  & = & a^2  (\partial_i (b^{(2)} + \Delta^{- 1} \partial^j \Xi_{0 j}) +
  \nu_i^{(2)} + (\delta_i^j - \partial_i \Delta^{- 1} \partial^j) \Xi_{0 j})
  \nonumber\\
  & \equiv & a^2  (\partial_i (b^{(2)} + \Delta^{- 1} \partial^j \Xi_{0 j}) +
  \nu_i^{(2)} +\mathcal{T}^j_i \Xi_{0 j}),  \label{56}\\
  \tilde{g}^{(2)}_{ij} & = & a^2  (- 2 \delta_{ij}  \tilde{\psi}^{(2)} + 2
  \partial_i \partial_j  \tilde{e}^{(2)} + \partial_i  \tilde{c}^{(2)}_j +
  \partial_j  \tilde{c}_i^{(2)} + \tilde{h}_{ij}^{(2)}) \nonumber\\
  & = & g_{ij}^{(2)} + 2\mathcal{L}_{\xi_1} g_{ij}^{(1)} +
  (\mathcal{L}_{\xi_2} +\mathcal{L}_{\xi_1}^2) g_{ij}^{(0)} \nonumber\\
  & = & a^2 \Big(- 2 \delta_{ij} \psi^{(2)} + 2 \partial_i \partial_j {e}^{(2)} +
  \partial_i {c}^{(2)}_j + \partial_j c_i^{(2)} + h_{ij}^{(2)} \nonumber\\
  &  & \left. + \frac{1}{2} \delta_{ij} (\delta^{kl} - \partial^k \Delta^{-
  1} \partial^l) \Xi_{kl} + \frac{1}{2} \partial_i \partial_j \Delta^{- 1} ((3
  \Delta^{- 1} \partial^k \partial^l - \delta^{kl}) \Xi_{kl}) \right.
  \nonumber\\
  &  & + \partial_j \Delta^{- 1} \partial^l  ((\delta^k_i - \partial_i
  \Delta^{- 1} \partial^k) \Xi_{kl}) + \partial_i \Delta^{- 1} \partial^k 
 ( (\delta^l_j - \partial_j \Delta^{- 1} \partial^l) \Xi_{kl}) \nonumber\\
  &  & + ( (\delta^k_i - \partial_i \Delta^{- 1} \partial^k) (\delta^l_j
  - \partial_j \Delta^{- 1} \partial^l) - \frac{1}{2} (\delta_{ij} -
  \partial_i \Delta^{- 1} \partial_j) (\delta^{kl} - \partial^k \Delta^{- 1}
  \partial^l) ) \Xi_{kl} \Big) \nonumber\\
  & = & a^2  \Big( - 2 \delta_{ij} \psi^{(2)} + 2 \partial_i \partial_j
  {e}^{(2)} + \partial_i {c}^{(2)}_j + \partial_j c_i^{(2)} + h_{ij}^{(2)} +
  \frac{1}{2} \delta_{ij} \mathcal{T}^{kl} \Xi_{kl}  + \partial_i
  \partial_j \Delta^{- 1}  (( \partial^k \Delta^{- 1} \partial^l -
  \frac{1}{2} \mathcal{T}^{kl} ) \Xi_{kl}) \nonumber\\
  &  &  + \partial_j \Delta^{- 1} \partial^l \mathcal{T}^k_i \Xi_{kl} +
  \partial_i \Delta^{- 1} \partial^k \mathcal{T}^l_j \Xi_{kl} + (
  \mathcal{T}^k_i \mathcal{T}^l_j - \frac{1}{2} \mathcal{T}_{ij}
  \mathcal{T}^{kl} ) \Xi_{kl} \Big),  \label{57}
\end{eqnarray}
\end{subequations}\vspace{5mm}\ruledown
\begin{multicols}{2}\hspace{-6.5mm}
where $\mathcal{T}^l_j \equiv \delta^l_j - \partial_j \Delta^{- 1} \partial^l$
is a transverse operator, and we define 
\begin{equation}
  \Xi_{\mu \nu} \equiv \frac{2\mathcal{L}_{\xi_1} g_{\mu \nu}^{(1)} +
  (\mathcal{L}_{\xi_2} +\mathcal{L}_{\xi_1}^2) g_{\mu \nu}^{(0)}}{a^2} .
\end{equation}
In Eqs.~(\ref{56}) and (\ref{57}), we have decomposed the gauge
transformation into the scalar, vector and tensor components. Therefore, the gauge
transformation of each component can be rewritten as
\begin{subequations}
\begin{eqnarray}
    \tilde{\phi}^{(2)} & = & \phi^{(2)} - \frac{1}{2} \Xi_{00},\\
    \tilde{b}^{(2)} & = & b^{(2)} + \Delta^{- 1} \partial^j \Xi_{0 j},\\
    \tilde{\nu}^{(2)}_i & = & \nu_i^{(2)} +\mathcal{T}^j_i \Xi_{0 j},\\
    \tilde{\psi}^{(2)} & = & \psi^{(2)} - \frac{1}{4} \mathcal{T}^{kl}
    \Xi_{kl},\\
    \tilde{e}^{(2)} & = & {e}^{(2)} + \frac{1}{2} \Delta^{- 1}  \Big(\big(
    \partial^k \Delta^{- 1} \partial^l - \frac{1}{2} \mathcal{T}^{kl} \big)
    \Xi_{kl}\Big),\\
    \tilde{c}_j^{(2)} & = & c_j^{(2)} + \Delta^{- 1} \partial^k
    \mathcal{T}^l_j \Xi_{kl},\\
    \tilde{h}_{ij}^{(2)} & = & h^{(2)}_{ij} + \left( \mathcal{T}^k_i
    \mathcal{T}^l_j - \frac{1}{2} \mathcal{T}_{ij} \mathcal{T}^{kl} \right)
    \Xi_{kl} .
\end{eqnarray}  \label{58}
\end{subequations}
In particular, the second order tensor perturbation is no longer invariant upon the gauge transformation. 
%It is shown to be mixed with the first order cosmological perturbations. 
Based on Eqs.~(\ref{10}), (\ref{12}) and (\ref{58}), we obtain the
gauge invariant second order variables as follows
\begin{subequations}
\begin{eqnarray}
    \Phi^{(2)} & = & \phi^{(2)} - \left( \partial_0 + \frac{\dot{a}}{a}
    \right) Y^0 +\mathcal{X}_{00},\\
    B^{(2)} & = & b^{(2)} - \partial_0 Y_S + Y^0 - \Delta^{- 1}
    \partial^j \mathcal{X}_{0 j},\\
    V_i^{(2)} & = & \nu_i^{(2)} - \delta_{ij} \partial_0 Y^j_T
    -\mathcal{T}^j_i \mathcal{X}_{0 j},\\
    \Psi^{(2)} & = & \psi^{(2)} + \frac{\dot{a}}{a} Y^0 + \frac{1}{4}
    \mathcal{T}^{kl} \mathcal{X}_{kl}\\
    E^{(2)} & = & {e}^{(2)} - Y_S - \frac{1}{2} \Delta^{- 1}  \Big(\big(
    \partial^k \Delta^{- 1} \partial^l - \frac{1}{2} \mathcal{T}^{kl} \big)
    \mathcal{X}_{kl}\Big),\\
    C_j^{(2)} & = & c_j^{(2)} - \delta_{jk} Y^k_T - \Delta^{- 1}
    \partial^k \mathcal{T}^l_j \mathcal{X}_{kl},\\
    H_{ij}^{(2)} & = & h_{ij}^{(2)} - \left( \mathcal{T}^k_i \mathcal{T}^l_j -
    \frac{1}{2} \mathcal{T}_{ij} \mathcal{T}^{kl} \right) \mathcal{X}_{kl},
\end{eqnarray} \label{59}
\end{subequations}
where we define
\begin{equation}
  \mathcal{X}_{\mu \nu} \equiv \frac{2\mathcal{L}_X g_{\mu \nu}^{(1)}
  -\mathcal{L}_X^2 g_{\mu \nu}^{(0)}}{a^2} . \label{67c}
\end{equation}
and we have decomposed $Y^i = : Y_T^i + \delta^{ij} \partial_j Y_S$, i.e., into the
transverse part and the longitudinal one. 
We find that $\mathcal{X}_{\mu\nu}$ depends on $X^{\mu}$ which determines the gauge invariant first order variables. 
In Appendix \ref{D}, we will present
an explicit expression of $\mathcal{X}_{\mu \nu}$. As shown in Eq.~(\ref{59}), all the gauge invariant variables are expressed in terms of $X^{\mu}$ and
$Y^{\mu}$, except that the gauge invariant second order tensor perturbation
$H^{(2)}_{ij}$ only depends on $X^{\mu}$. %And differed from the first-order tensor perturbations $H_{ij}^{(1)}$, 
%The setup of $H^{(2)}_{ij} = 0$ leads to an additional
%constraint for $X^{\mu}$ and seems to be unphysical. 
%This implies that
%$H^{(2)}_{ij}$ should exist, if one wishes to study evolution of second order
%cosmological perturbations in a gauge invariant scenario.
As the explicit expression of $X^{\mu}$ has been known, only $Y^{\mu}$ is undetermined in Eq.~(\ref{59}). 
Therefore, we will show how to express $Y^{\mu}$
in terms of the first and second order metric perturbations in the following.

\subsection{Gauge invariant second order Newtonian variables}\label{IV.A}

To obtain the gauge invariant Newtonian variables, we derive the expression of $Y^\mu$ from $B^{(2)}=E^{(2)}=C_j^{(2)}=0$. 
\iffalse
\begin{eqnarray}
  0 =B^{(2)} & = & b^{(2)} - \Delta^{- 1} \partial^j \mathcal{X}_{0
  j} + \partial_0 Y_S + Y^0,  \label{61}\\
  0 =E^{(2)} & = & {e}^{(2)} - \frac{1}{2} \Delta^{- 1}  \left(
  \partial^k \Delta^{- 1} \partial^l - \frac{1}{2} \mathcal{T}^{kl} \right)
  \mathcal{X}_{kl} - Y_S, \\
  0 =C_j^{(2)} & = & c_j^{(2)} - \Delta^{- 1} \partial^k
  \mathcal{T}^l_j \mathcal{X}_{kl} - \delta_{jk} Y^k_T .  \label{63}
\end{eqnarray}
\fi
The explicit expression of $Y^{\mu}$ is given by
\begin{subequations}
\begin{eqnarray}
  Y^0 & = & \partial_0 {e}^{(2)} - b^{(2)} + \Delta^{- 1} \partial^j
  \mathcal{X}_{0 j} - \frac{1}{2}   ( \partial^k \Delta^{- 2}
  \partial^l \partial_0 \nonumber\\
  & &- \frac{1}{2} \Delta^{- 1}\mathcal{T}^{kl}\partial_0 ) 
  \mathcal{X}_{kl},  \label{64}\\
  Y^i & = & \delta^{ij}  (\partial_j {e}^{(2)} + c_j^{(2)}) + ( \frac{1}{4}
    \partial^i\partial^k \Delta^{- 2} \partial^l \nonumber\\
    & &+ \frac{1}{4}\delta^{kl}  \partial^i \Delta^{- 1} -
  \delta^{li} \Delta^{- 1} \partial^k ) \mathcal{X}_{kl} .  \label{65}
\end{eqnarray}
\end{subequations}
We find that $Y^{\mu}$ depends on a choice of the first order
variable $X^{\mu}$.
Therefore, the gauge invariant second order metric perturbations turn to be
\begin{subequations}
\begin{eqnarray}
    g^{(\mathrm{\rm{GI}, 2)}}_{00} & = & - 2 a^2 \Phi^{(2)},\\
    g_{0 i}^{(\mathrm{\rm{GI}, 2)}} & = & a^2 V_i^{(2)},\\
    g_{ij}^{(\mathrm{\rm{GI}, 2)}} & = & - 2 a^2 \delta_{ij} \Psi^{(2)} +
    a^2 H_{ij}^{(2)},
\end{eqnarray}\label{66}
\end{subequations}
where the gauge invariant variables are defined as
\begin{subequations}
\begin{eqnarray}
  \Phi^{(2)} & = & \phi^{(2)} - ( \frac{\dot{a}}{a} +  \partial_0  ) 
  (\partial_0 {e}^{(2)} - b^{(2)}) +\mathcal{X}_{00} \nonumber\\
  & & - (\frac{\dot{a}}{a} + \partial_0 )  (\Delta^{- 1} \partial^j \mathcal{X}_{0 j} -
 3 \Delta^{- 2} \partial^k \partial^l \mathcal{X}_{kl} \nonumber\\
 & & + \delta^{kl}   \Delta^{- 1}\mathcal{X}_{kl}),  \label{74c}\\
  \Psi^{(2)} & = & \psi^{(2)} + \frac{\dot{a}}{a}  (\partial_0 {e}^{(2)} -
  b^{(2)}) \nonumber\\
  & &+ \frac{\dot{a}}{a}  (\Delta^{- 1} \partial^j \mathcal{X}_{0 j} -
 3 \Delta^{- 2} \partial^k \partial^l \mathcal{X}_{kl} \nonumber\\
 & & +   \delta^{kl}  \Delta^{- 1}\mathcal{X}_{kl}) +\mathcal{T}^{kl} \mathcal{X}_{kl}, \\
  V_i^{(2)} & = & \nu_i^{(2)} - \partial_0 c_i^{(2)} + \Delta^{- 1} \partial^k
  \mathcal{T}^l_i \partial_0 \mathcal{X}_{kl} \nonumber\\
  & & -\mathcal{T}^j_i \mathcal{X}_{0  j}, \\
  H_{ij}^{(2)} & = & h^{(2)}_{ij} - ( \mathcal{T}^k_i \mathcal{T}^l_j -
  \frac{1}{2} \mathcal{T}_{ij} \mathcal{T}^{kl} ) \mathcal{X}_{kl} . 
  \label{67}
\end{eqnarray}
\end{subequations}
%The gauge invariant second order
%variables can not be completely determined before the $X^{\mu}$ are given. One the other side as shown in Eqs.~(\ref{74c})--(\ref{67}), there are also not any constraints on the
%$\mathcal{X}_{\mu \nu}$.  In
%this work, we show a more compacted form of the gauge invariant metric
%perturbations than that in previous work
%{\cite{christopherson_comparing_2011}}\cite{malik_cosmological_2009}.

\subsection{Gauge invariant second order synchronous variables}\label{IV.B}
For gauge invariant synchronous  variables, the $Y^\mu$ is determined by making use of $\Phi^{(2)} =0$, $B^{(2)} = 0 $ and $ V^{(2)}_i = 0$. The explicit expression of the $Y^\mu$ takes the form of
\iffalse
\begin{eqnarray}
  0 = \Phi^{(2)} & = & \phi^{(2)} + \frac{1}{2} \mathcal{X}_{00} - \frac{1}{a}
  \partial_0  (aY^0), \\
  0 =B^{(2)} & = & b^{(2)} - \Delta^{- 1} \partial^j \mathcal{X}_{0
  j} - \partial_0 Y_S + Y^0, \\
  0 = V^{(2)}_i & = & \nu_i^{(2)} -\mathcal{T}^j_i \mathcal{X}_{0 j} -
  \delta_{ki} \partial_0 Y_T^k . 
\end{eqnarray}
It leads to
\fi
\begin{subequations}
\begin{eqnarray}
  Y^0 & = & \frac{1}{a}  \int \mathrm{d} \eta \{ a \phi^{(2)} +
  \frac{1}{2} a\mathcal{X}_{00} \},  \label{81c}\\
  Y^k & = & \delta^{ki}  \int \mathrm{d} \eta \{ \nu_i^{(2)} + \partial_i
  b^{(2)} + \frac{1}{a}  \int \mathrm{d} \eta \{a \partial_i \phi^{(2)} \}
  \} \nonumber\\
   & &- \int \mathrm{d} \eta \{ \delta^{kj} \mathcal{X}_{0 j} -
  \frac{1}{2 a}  \int \mathrm{d} \eta \{a \partial^k \mathcal{X}_{00} \}
  \} .  \label{82c}
\end{eqnarray}
\end{subequations}
Therefore, we have the gauge invariant second order metric perturbations in
the form of
\begin{subequations}
\begin{eqnarray}
    g_{00}^{(\mathrm{\rm{GI}, 2)}} & = & 0,\\
    g_{0 i}^{(\mathrm{\rm{GI}, 2)}} & = & 0,\\
    g_{ij}^{(\mathrm{\rm{GI}, 2)}} & = & a^2  (- 2 \Psi^{(2)} \delta_{ij} +
    2 \partial_i \partial_j E^{(2)} \nonumber\\
    & & + \partial_i C_j^{(2)} +
    \partial_j C_i^{(2)} + H_{ij}^{(2)}),
\end{eqnarray}
\end{subequations}
where the gauge invariant variables are defined with
\begin{subequations}
\begin{eqnarray}
  \Psi^{(2)} & = & \psi^{(2)} + \frac{\dot{a}}{a^2}  \int \mathrm{d} \eta
  \{ a \phi^{(2)} + \frac{1}{2} a\mathcal{X}_{00} \}
  +\mathcal{T}^{kl} \mathcal{X}_{kl},  \label{84c}\\
  E^{(2)} & = & {e}^{(2)} - \int \mathrm{d} \eta \{ b^{(2)} +
  \frac{1}{a}  \int \mathrm{d} \eta' \{a \phi^{(2)} \} \} \nonumber\\
  &  & + \int \mathrm{d} \eta \{ \Delta^{- 1} \partial^j \mathcal{X}_{0
  j} - \frac{1}{2 a}  \int \mathrm{d} \eta' \{a\mathcal{X}_{00} \} \} \nonumber\\
& & -
  \frac{1}{4}   (3 \Delta^{- 2} \partial^k \partial^l - 
  \delta^{kl}\Delta^{- 1}) \mathcal{X}_{kl}, \\
  C_j^{(2)} & = & c_j^{(2)} - \int \nu_j^{(2)} \mathrm{d} \eta
  +\mathcal{T}^k_j  \int \mathcal{X}_{0 k} \mathrm{d} \eta \nonumber\\
  & & - \Delta^{- 1}
  \partial^k \mathcal{T}^l_j \mathcal{X}_{kl}, \\
  H_{ij}^{(2)} & = & h^{(2)}_{ij} - ( \mathcal{T}^k_i \mathcal{T}^l_j -
  \frac{1}{2} \mathcal{T}_{ij} \mathcal{T}^{kl} ) \mathcal{X}_{kl} . 
  \label{90}
\end{eqnarray}
\end{subequations}
The gauge invariant synchronous variables are
different from the gauge invariant Newtonian variables, except the tensor
perturbations. In both of the two cases, $H_{ij}^{(2)}$ is completely
determined by the choice of the gauge invariant first order variables. 

\subsection{Conversion among different families of gauge invariant second order variables}

Compared with the first order case in the previous section, a conversion between two different families of gauge invariant second order variables is more complicated, since the gauge invariant second order variables are also dependent of the choice of the gauge invariant first order variables. 
To be specific, for two different families of gauge invariant second order metric
perturbations $g_{\mu \nu}^{(\mathrm{\rm{GI}, A, 2)}}$ and $g_{\mu
\nu}^{(\mathrm{\rm{GI}, B, 2)}}$, the conversion between them can be derived
from
\begin{eqnarray}
  g_{\mu \nu}^{(\mathrm{\rm{GI}, A, 2)}} - g_{\mu \nu}^{(\mathrm{\rm{GI},
  B, 2)}} & = & g_{\mu \nu}^{(2)} - 2\mathcal{L}_{X^A} g_{\mu \nu}^{(1)} -
  (\mathcal{L}_{Y^A} -\mathcal{L}_{X^A}^2) g_{\mu \nu}^{(0)} \nonumber\\
  &  & - (g_{\mu
  \nu}^{(2)} - 2\mathcal{L}_{X^B} g_{\mu \nu}^{(1)} - (\mathcal{L}_{Y^B}
  -\mathcal{L}_{X^B}^2) g_{\mu \nu}^{(0)}) \nonumber\\
  & = & 2\mathcal{L}_{(X^B - X^A)}  (g_{\mu \nu}^{(\mathrm{\rm{GI}, B, 1)}}
  +\mathcal{L}_{X^B} g_{\mu \nu}^{(0)}) \nonumber\\
  &  &+ (\mathcal{L}_{(Y^B - Y^A)}
  -\mathcal{L}_{X^B}^2 -\mathcal{L}_{X^A}^2) g_{\mu \nu}^{(0)} \nonumber\\
  & = & - 2\mathcal{L}_{(X^A - X^B)} g_{\mu \nu}^{(\mathrm{\rm{GI}, B, 1)}} \nonumber\\
  &  &-
  (\mathcal{L}_{(Y^A - Y^B + [X^A, X^B])}\nonumber\\
  &  & -\mathcal{L}_{(X^A - X^B)}^2) g_{\mu
  \nu}^{(0)} .  \label{94}
\end{eqnarray}
If we let $Z_2^{\mathrm{AB}} \equiv Y^A - Y^B + [X^A, X^B]$ and
$Z_1^{\mathrm{AB}} \equiv X^A - X^B$, Eq.~(\ref{94}) can be rewritten as
\begin{equation}\label{sai:1}
  g_{\mu \nu}^{(\mathrm{\rm{GI}, A, 2)}} = g_{\mu \nu}^{(\mathrm{\rm{GI},
  B, 2)}} - 2\mathcal{L}_{Z^{\mathrm{AB}}_1} g_{\mu \nu}^{(\mathrm{\rm{GI},
  B, 1)}} - (\mathcal{L}_{Z_2^{\mathrm{AB}}}
  -\mathcal{L}_{Z^{\mathrm{AB}}_1}^2) g_{\mu \nu}^{(0)} .
\end{equation}
This conversion was also mentioned by Nakamura
\cite{Nakamura:2019zbe} in a different formula. One can easily check that both $Z_1^{\mathrm{AB}}$ and $Z_2^{\mathrm{AB}}$ are
gauge invariant by making use of Eqs.~(\ref{13}) and (\ref{14}). 
The above formula is generic. 
It is obvious that the gauge invariant second order variables depend on the gauge invariant first order variables, since in general we have $Z_{1}^{\mathrm{AB}}\neq 0$, namely, two different families of gauge invariant first order variables. 
This generic case will be studied later in this section. 
When we take the same family of the gauge invariant variables at first order, i.e., $Z_{1}^{\mathrm{AB}}=0$, Eq.~(\ref{sai:1}) can be reduced to a simpler form $g_{\mu \nu}^{(\mathrm{\rm{GI}, A, 2)}} = g_{\mu \nu}^{(\mathrm{\rm{GI},
  B, 2)}}-\mathcal{L}_{Z_2^{\mathrm{AB}}}g_{\mu \nu}^{(0)}$, where $Z_2^{\mathrm{AB}}= Y^A - Y^B$. 
For this simple case, the formula is as similar as that of Eq.~(\ref{59a}). 
%This implies that the conversion depends on the choice of the second order 

We also have infinite families of gauge invariant second order variables.
Similar to $Z_1^{\mathrm{AB}}$ in Eq.~(\ref{60}), the infinitesimal vector 
$Z_2^{\mathrm{AB}}$ can also be expressed as a linear combination of the
gauge invariant second order variables $\mathcal{A}^{(2)}$, $\mathcal{B}^{(2)}$,
$\mathcal{C}^{(2)}_i$ and $\mathcal{D}_{\sigma \kappa}^{(2)}$,
\begin{subequations}
\begin{eqnarray}
  Z_2^{0, \mathrm{AB}} & = & \hat{M}_1^0 \mathcal{A}^{(2)} + \hat{M}_2^0
  \mathcal{B}^{(2)} + \hat{M}_3^{0 i} \mathcal{C}_i^{(2)} \nonumber\\
  &  &+ \hat{M}^{0, \sigma
  \kappa}_4 \mathcal{D}_{\sigma \kappa}^{(2)},  \label{76-1}\\
  Z_2^{k, \mathrm{AB}} & = & \hat{M}_1^k \mathcal{A}^{(2)} + \hat{M}_2^k
  \mathcal{B}^{(2)} + \hat{M}_3^{ki} \mathcal{C}_i^{(2)} \nonumber\\
  &  & + \hat{M}^{k, \sigma
  \kappa}_4 \mathcal{D}_{\sigma \kappa}^{(2)},  \label{76}
\end{eqnarray}
\end{subequations}
where $\hat{M}_1^\mu$, $\hat{M}_2^\mu$, $\hat{M}_3^{\mu i}$ and $\hat{M}^{\mu, \sigma
\kappa}_4$ are four arbitrary linear operators that are irrelative to any
perturbations, and the gauge-invariant variables are defined as
\begin{subequations}
\begin{eqnarray}
  \mathcal{A}^{(2)} & \equiv & \partial_0  ( \frac{a^2}{\dot{a}}
  \psi^{(2)} ) + a \phi^{(2)} + \frac{1}{4} \partial_0  (
  \frac{a^2}{\dot{a}} \mathcal{T}^{kl} \mathcal{X}_{kl} ) \nonumber\\
  & & + \frac{1}{2}
  a\mathcal{X}_{00}, \\
  \mathcal{B}^{(2)} & \equiv & \partial_0 {e}^{(2)} - b^{(2)} +
  \frac{a}{\dot{a}} \psi^{(2)} + \Delta^{- 1} \partial^j \mathcal{X}_{0 j} \nonumber\\
  & & +
  \frac{1}{4}  ( \frac{a}{\dot{a}} \mathcal{T}^{kl} -  3
  \Delta^{- 2} \partial^k \partial^l \partial_0 + \Delta^{- 1}\delta^{kl}\partial_0  )
  \mathcal{X}_{kl}, \\
  \mathcal{C}_k^{(2)} & \equiv & \nu_k^{(2)} - \partial_0 c_k^{(2)} +
  \Delta^{- 1} \partial^i \mathcal{T}^l_k \partial_0 \mathcal{X}_{il}
  -\mathcal{T}^l_k \mathcal{X}_{0 l}, \\
  \mathcal{D}_{\mu \nu}^{(2)} & \equiv &
  \frac{1}{a^2}{(2\mathcal{L}_{Z_1^{\mathrm{AB}}} g_{\mu \nu}^{(\mathrm{\rm{GI}, B,
  1)}} -\mathcal{L}_{Z_1^{\mathrm{AB}}}^2 g_{\mu \nu}^{(0)})} . 
\end{eqnarray}
\end{subequations}
Here, we express $\mathcal{A}^{(2)}$, $\mathcal{B}^{(2)}$ and $\mathcal{C}^{(2)}_i$ in terms of $\mathcal{X}_{\mu \nu}$, which is completely determined by $X^{B}$. 
%In fact, we can obtain $\mathcal{A}^{(2)}$, $\mathcal{B}^{(2)}$ and $\mathcal{C}^{(2)}_i$ by making use of Eq.~(\ref{58}). Here, 
%$\mathcal{D}_{\sigma \kappa}^{(2)}$ only consists of the square of the
%gauge invariant first order variables. From 
Based on Eqs.~(\ref{76-1}) and (\ref{76}), we can also obtain a new family of the gauge invariant second order variables by a conversion from a given family of the gauge invariant second order variables. %via the linear operators $\hat{M}_1^\mu$, $\hat{M}_2^\mu$, $\hat{M}_3^{\mu i}$ and $\hat{M}^{\mu, \sigma 	\kappa}_4$.
This prediction is similar to that of the first order case. 

We consider three typical cases in the following.  
In the case that $g_{\mu \nu}^{(\mathrm{\rm{GI}, A,n)}}$ and $g_{\mu
\nu}^{(\mathrm{\rm{GI}, B,n)}}$ ($n=1,2$) are synchronous and Newtonian, respectively, we find that $Z_1^{\mu,
\mathrm{AB}}$ takes the form of Eqs.~(\ref{59b}) and (\ref{60b}), and $Z_2^{\mu, \mathrm{AB}}$ is shown to be
\begin{subequations}
\begin{eqnarray}
  Z^{0, \mathrm{AB}}_2 & = & \frac{1}{a}  \int \mathcal{A}^{(2)} \mathrm{d}
  \eta -\mathcal{B}^{(2)} + \frac{1}{2 a}  \int \mathrm{d} \eta
  \{a\mathcal{D}_{00}^{(2)} \}, \\
  Z_2^{j, \mathrm{AB}} & = & \delta^{jk} \partial_k  \int \mathrm{d} \eta
  \{ \frac{1}{a}  \int \mathcal{A}^{(2)} \mathrm{d} \eta' \} \nonumber\\
  & & -
  \delta^{jk} \partial_k  \int \mathcal{B}^{(2)} \mathrm{d} \eta + \delta^{j k}
  \int \mathcal{C}_k^{(2)} \mathrm{d} \eta \nonumber\\
  &  & - \int \mathrm{d} \eta \big\{ \delta^{j k} \mathcal{D}_{0 k}^{(2)} -
  \frac{1}{2 a}  \int \mathrm{d} \eta \{a \partial^j \mathcal{D}_{00}^{(2)} \}
  \big\} . 
\end{eqnarray}
\end{subequations}
Since the vectors $X^{\mu}$ and $Y^{\mu}$ are independent,
we can choose, e.g., the gauge invariant Newtonian variables at the
first order and the gauge invariant synchronous variables at the second order, and vice versa. 
First,   in the case  that $g_{\mu \nu}^{(\mathrm{\rm{GI}, A,1)}}$ and $g_{\mu
	\nu}^{(\mathrm{\rm{GI}, B,1)}}$ are Newtonian while $g_{\mu \nu}^{(\mathrm{\rm{GI}, A,2)}}$ and $g_{\mu
	\nu}^{(\mathrm{\rm{GI}, B,2)}}$ are synchronous and Newtonian, respectively, we obtain $Z^{\mu, \mathrm{AB}}_1 = 0$ and 
%as has been mentioned above, we consider the simple case $X^A = X^B$   but $Y^A \neq Y^B$, implying that we choose  gauge invariant Newtonian variables at first order but different at second order. 
\begin{subequations}
\begin{eqnarray}
  Z^{0, \mathrm{AB}}_2 & = & \frac{1}{a}  \int \mathcal{A}^{(2)} \mathrm{d}
\eta -\mathcal{B}^{(2)} , \\
Z_2^{j, \mathrm{AB}} & = & \delta^{jk} \partial_k  \int \mathrm{d} \eta
\{ \frac{1}{a}  \int \mathcal{A}^{(2)} \mathrm{d} \eta' \} \nonumber\\
& & -
\delta^{jk} \partial_k  \int \mathcal{B}^{(2)} \mathrm{d} \eta + \delta^{jk}
\int \mathcal{C}_k^{(2)} \mathrm{d} \eta . 
\end{eqnarray}
\end{subequations}
Second, in the case that $g_{\mu \nu}^{(\mathrm{\rm{GI}, A,1)}}$ and $g_{\mu
	\nu}^{(\mathrm{\rm{GI}, B,1)}}$ are synchronous and Newtonian, respectively, while both $g_{\mu \nu}^{(\mathrm{\rm{GI}, A,2)}}$ and $g_{\mu	\nu}^{(\mathrm{\rm{GI}, B,2)}}$ are Newtonian, we find that $Z_1^{\mu,  \rm AB}$ takes the same form as Eqs.~(\ref{59b}) and (\ref{60b}), and $Z_2^{\mu, \mathrm{AB}}$ turns to be
\begin{subequations}
\begin{eqnarray}
Z^{0,\rm{AB}}_2 & = &  \Delta^{- 1} \partial^j
\mathcal{D}_{0 j}^{(2)} \nonumber\\
& & - \frac{1}{2}   ( \partial^k \Delta^{- 2}
\partial^l\partial_0 - \frac{1}{2} \Delta^{- 1}\mathcal{T}^{kl}\partial_0 ) 
\mathcal{D}_{kl}^{(2)},  \\
Z^{i, \rm{AB}}_2 & = & ( \frac{1}{4}
 \partial^i\partial^k \Delta^{- 2} \partial^l + \frac{1}{4}\delta^{kl} \Delta^{- 1}\partial^i \nonumber\\
 & & -
\delta^{li} \Delta^{- 1} \partial^k ) \mathcal{D}_{kl}^{(2)} ,
\end{eqnarray}
\end{subequations}
which is expressed in terms of the square of the first order metric perturbations only.
The above two cases are called the gauge invariant hybrid variables. 
%Compared with the case at the first order [Eqs.~(\ref{59b}) and (\ref{60b})], there
%are additional terms of the $D_{\sigma \kappa}^{(2)}$. 

\section{Gauge invariant equations of motion for the second order cosmological perturbations}\label{V}

In this section, we will derive the equations of motion of the second order cosmological perturbations, which are sourced from the first order scalar perturbations in the gauge invariant framework. 
For simplicity, we will consider the gauge invariant Newtonian, synchronous, and hybrid
variables that have been introduced in the previous sections. 

\subsection{Gauge invariant energy-momentum tensor up to second order}
On the side of matter, we expand the energy-momentum tensor of the perfect fluid up
to second order, i.e.,
\begin{eqnarray}
  T_{\mu \nu}^{(\mathrm{\rm{GI})}} & \equiv & T_{\mu \nu}^{(0)} + T_{\mu
  \nu}^{(\mathrm{\rm{GI}, 1)}} + \frac{1}{2} T_{\mu
  \nu}^{(\mathrm{\rm{GI}, 2)}} +\mathcal{O} (T_{\mu
  \nu}^{(\mathrm{\rm{GI}, 3)}}), 
\end{eqnarray}
where
\end{multicols}
\hspace{0mm}\vspace{5mm}\ruleup
\begin{subequations}
\begin{eqnarray}
  T^{(0)}_{\mu \nu} & = & u_{\mu}^{(0)} u_{\nu}^{(0)}  (\rho^{(0)} + P^{(0)})
  + g_{\mu \nu}^{(0)} P^{(0)},  \label{83}\\
  T^{(\mathrm{\rm{GI}, 1)}}_{\mu \nu} & = & u_{\mu}^{(\mathrm{\rm{GI},
  1)}} u_{\nu}^{(0)}  (\rho^{(0)} + P^{(0)}) + u^{(0)}_{\mu}
  u^{(\mathrm{\rm{GI}, 1)}}_{\nu}  (\rho^{(0)} + P^{(0)}) + u_{\mu}^{(0)}
  u_{\nu}^{(0)}  (\rho^{(\mathrm{\rm{GI}, 1)}} + P^{(\mathrm{\rm{GI},
  1)}}) \nonumber\\
  &  & + g_{\mu \nu}^{(0)} P^{(\mathrm{\rm{GI}, 1)}} + g_{\mu
  \nu}^{(\mathrm{\rm{GI}, 1)}} P^{(0)},  \label{84}\\
  T^{(\rm{GI}, 2)}_{\mu \nu} & = & u_{\mu}^{(\rm{GI}, 1)}
  u_{\nu}^{(\rm{GI}, 1)} (\rho^{(0)} + P^{(0)}) + u_{\mu}^{(\rm{GI}, 1)}
  u_{\nu}^{(0)} (\rho^{(\rm{GI}, 1)} + P^{(\rm{GI}, 1)}) + u_{\mu}^{(0)}
  u_{\nu}^{(\rm{GI}, 1)} (\rho^{(\rm{GI}, 1)} + P^{(\rm{GI}, 1)})
  \nonumber\\
  &  & u_{\mu}^{(\mathrm{\rm{GI}, 2)}} u_{\nu}^{(0)}  (\rho^{(0)} +
  P^{(0)}) + u^{(0)}_{\mu} u^{(\mathrm{\rm{GI}, 2)}}_{\nu}  (\rho^{(0)} +
  P^{(0)}) + u_{\mu}^{(0)} u_{\nu}^{(0)}  (\rho^{(\mathrm{\rm{GI}, 2)}} +
  P^{(\mathrm{\rm{GI}, 2)}}) \nonumber\\
  &  & + g_{\mu \nu}^{(\mathrm{\rm{GI}, 1)}} P^{(\mathrm{\rm{GI}, 1)}} +
  g_{\mu \nu}^{(0)} P^{(\mathrm{\rm{GI}, 2)}} + g_{\mu
  \nu}^{(\mathrm{\rm{GI}, 2)}} P^{(0)} . 
\end{eqnarray}
\end{subequations}
\vspace{5mm}\ruledown
\begin{multicols}{2}\hspace{-6.5mm}
Here, $\rho^{(0)}, P^{(0)}$ denote the background density and pressure, respectively.
The $\rho^{(\mathrm{\rm{GI}, n)}}$,
$u_{\mu}^{(\mathrm{\rm{GI}, n)}}$ and $P^{(\mathrm{\rm{GI},
n)}}$ denote the gauge invariant $n$-th order density, pressure and
velocity perturbations, respectively. As has been suggested in Eqs.~(\ref{15-2}) and (\ref{15-3}), the
gauge invariant matter perturbations are formulated as
\begin{subequations}
\begin{eqnarray}
    \rho^{(\mathrm{\rm{GI}, 1)}} & = & \rho^{(1)} -\mathcal{L}_X
    \rho^{(0)},\\
    P^{(\mathrm{\rm{GI}, 1)}} & = & P^{(1)} -\mathcal{L}_X P^{(0)},\\
    u_{\mu}^{(\mathrm{\rm{GI}, 1)}} & = & u_{\mu}^{(1)} -\mathcal{L}_X
    u_{\mu}^{(0)},\\
    \rho^{(\mathrm{\rm{GI}, 2)}} & = & \rho^{(2)} - 2\mathcal{L}_X
    \rho^{(1)} - (\mathcal{L}_Y -\mathcal{L}_X^2) P^{(0)},\\
    P^{(\mathrm{\rm{GI}, 2)}} & = & P^{(2)} - 2\mathcal{L}_X P^{(1)} -
    (\mathcal{L}_Y -\mathcal{L}_X^2) P^{(0)},\\
    u_{\mu}^{(\mathrm{\rm{GI}, 2)}} & = & u^{(2)}_{\mu} - 2\mathcal{L}_X
    u_{\mu}^{(1)} - (\mathcal{L}_Y -\mathcal{L}_X^2) u^{(0)}_{\mu},
\end{eqnarray} \label{87}
\end{subequations}
%where expressions of variables $X^{\mu}$ and $Y^{\mu}$ have been shown in
%Eqs.~(\ref{31}), (\ref{39}), (\ref{64}), (\ref{65}), (\ref{81c}) and
%(\ref{82c}). 
%As is suggested in Appendix~\ref{B-1},
%one can reproduce Eqs.~(\ref{15}) and (\ref{15-1}) by making use of Eq.~(\ref{87}) . In order to evaluate Einstein field equations using $xPand$ \cite{pitrou_xpand_2013}, t
The velocity
field of the perfect fluid can be redefined as
%\begin{equation}
  $(\upsilon^i)^{(\mathrm{\rm{GI}, n)}} \equiv a
  (u^i)^{(\mathrm{\rm{GI}, n)}}$,
%\end{equation}
where $(u^0)^{(\mathrm{\rm{GI}, n)}}$
could be determined via $g_{\mu \nu} u^{\mu} u^{\nu} = - 1$. 
The equation of state $w$ and the speed of sound $c_s$ (and $c_{s}^{(2)}$) are defined as
\begin{subequations}
\begin{eqnarray}
    P^{(0)} & = & w \rho^{(0)},\\
    P^{(\mathrm{\rm{GI}},1)} & = & c_s^2 \rho^{(\mathrm{\rm{GI}},1)} ,\\
    P^{(\mathrm{\rm{GI}},2)} & = & (c_s^{(2)})^2 \rho^{(\mathrm{\rm{GI}},2)}.
\end{eqnarray}\label{89-1}
\end{subequations}
All of them are gauge invariant. 
In principle, one could choose $c_{s}^{(2)}$ to be equal to $c_{s}$ for the adiabatic perturbation with a constant speed of sound \cite{Inomata:2020cck}. 
%The $w = 0$ and $w = \frac{1}{3}$ could describe matter dominated and
%radiation dominated epochs of the Universe, respectively. From the
%Eq.~(\ref{89-1}), it indicates that the $w$, $c_s$ and $c_s^{(2)}$ are all the
%zeroth-order variables and should be gauge invariant.

\subsection{Second order cosmological perturbations induced by the first order
Newtonian variables}

As a first step, we study the gauge invariant Newtonian metric perturbations at first order and their equations of motion. 
The gauge invariant metric up to first order is given by
\begin{eqnarray}
  g_{\mu \nu}^{(\mathrm{\rm{GI})}} \mathrm{d} x^{\mu} \mathrm{d} x^{\nu} & =
  & - a^2  (1 + 2 \Phi^{(1)}) \mathrm{d} \eta^2 + 2 a^2 V_i^{(1)} \mathrm{d}
  \eta \mathrm{d} x^i \nonumber\\
  & & + a^2 \delta_{ij} (1 - 2 \Psi^{(1)}) \mathrm{d} x^i
  \mathrm{d} x^j,  \label{81}
\end{eqnarray}
where the gauge invariant first order Newtonian variables have been shown in
Eq.~(\ref{31-1})--(\ref{31-3}). Here, we neglect the first order tensor perturbation, i.e., $H^{(1)}_{ij} = 0$. 
Differed from the neglect of the scalar or vector perturbations, the neglect of $H^{(1)}_{ij}$ does not violate the gauge invariance that is defined with all the diffeomorphisms (namely, arbitrary $\xi_1^{\mu}$). 
At second order, we will consider that
the gauge invariant metric perturbations are Newtonian and
synchronous, respectively.

Based on Eqs.~(\ref{16})
and (\ref{17}), we express the temporal derivative of conformal Hubble parameter, the density and velocity perturbations in terms of the gauge invariant first order metric perturbations, i.e., 
\begin{subequations}
\begin{eqnarray}
  \dot{\mathcal{H}} & = & - \frac{1}{2}  (1 + 3 w) \mathcal{H}^2, 
  \label{91-0}\\
  \rho^{(0)} & = & \frac{3\mathcal{H}^2}{\kappa a^2},  \label{91}\\
  \rho^{(\mathrm{\rm{GI}, 1)}} & = & \frac{- 6\mathcal{H}^2 \Phi^{(1)} -
  6\mathcal{H} \partial_0 \Psi^{(1)} + 2 \Delta \Psi^{(1)}}{\kappa a^2}, \\
  \upsilon_i^{(\mathrm{\rm{GI}, 1)}} & = & - V_i^{(1)} + \frac{\Delta
  V_i^{(1)} - 4 \partial_i  (\mathcal{H} \Phi^{(1)} + 4 \partial_0
  \Psi^{(1)})}{6 (1 + w) \mathcal{H}^2}, 
\end{eqnarray}
\end{subequations}
where $\upsilon_i^{(\mathrm{\rm{GI}, 1)}} = \delta_{ij}
(\upsilon^i)^{(\mathrm{\rm{GI}, 1)}}$. 
By substituting the above equations into the spatial part of the
first order Einstein field equation, we obtain 
\begin{subequations}
\begin{eqnarray}
  \Phi^{(1)} & = & \Psi^{(1)}, \\
  V^{(1)}_i & = & 0,  \label{95}
\end{eqnarray}
\end{subequations}
%where we disregard the decaying mode of the vector perturbations. 
and the equation of motion of $\Phi^{(1)}$, i.e, 
\begin{equation}
  \partial_0^2 \Phi^{(1)} + 3 (1 + c_s^2) \mathcal{H} \partial_0 \Phi^{(1)} +
  3 (c^2_s - w) \mathcal{H}^2 \Phi^{(1)} - c_s^2 \Delta \Phi^{(1)} = 0.
\end{equation}
Here, we disregard the decaying mode of the first order vector perturbations.
% For matter perturbations, we disregard the anisotropic stress, since the impact of it is expected to be small  \cite{Saga:2014jca}.
%These results are well-known in previous works \cite{mukhanov_theory_1992}
In fact, the above results about the first order cosmological perturbations are well known \cite{Mukhanov:1990me}. 

\subsubsection{Gauge invariant second order Newtonian variables}\label{sub.1}

At second order, the gauge invariant Newtonian metric perturbations are given by
\begin{eqnarray}
  g_{\mu \nu}^{(\mathrm{\rm{GI}, 2)}} \mathrm{d} x^{\mu} \mathrm{d} x^{\nu}
  & = & 2 a^2 \Phi^{(2)} \mathrm{d} \eta^2 + 2 a^2 V_i^{(2)} \mathrm{d} \eta
  \mathrm{d} x^i \nonumber\\
  & & + a^2  (- 2 \delta_{ij} \Psi^{(2)} + H_{ij}^{(2)}) \mathrm{d}
  x^i \mathrm{d} x^j. 
\end{eqnarray}
The explicit expressions of the gauge invariant second order variables
$\Phi^{(2)}$, $\Psi^{(2)}$, $V_i^{(2)}$ and $H_{ij}^{(2)}$ are presented in
Eqs.~(\ref{74c})--(\ref{67}) with the expression of $\mathcal{X}_{\mu \nu}^N$
in Appendix \ref{D}. Based on the temporal components of the second order Einstein
field equations and Eqs.~(\ref{91-0})--(\ref{95}), we can express the
gauge invariant second order density perturbations in terms of the gauge invariant
metric perturbations, i.e.,
\end{multicols}
\hspace{0mm}\vspace{5mm}\ruleup
\begin{eqnarray}\label{eq:rhogi2}
  \rho^{(\rm{GI}, 2)} & = & \frac{2}{3 (1 + w) \kappa a^2 \mathcal{H}^2} (9
  (1 + w) \mathcal{H}^4 (4 (\Phi^{(1)})^2 - \Phi^{(2)}) - 9 (1 + w)
  \mathcal{H}^3 \partial_0 \Psi^{(2)}  \nonumber\\
  &  & +\mathcal{H}^2 (9 (1 + w) (\partial_0 \Phi^{(1)})^2 + 24 (1 + w)
  \Phi^{(1)} \Delta \Phi^{(1)} + 3 (1 + w) \Delta \Psi^{(2)} + (5 + 9 w)
  \partial_i \Phi^{(1)} \partial^i \Phi^{(1)}) \nonumber\\
  &  &  - 8\mathcal{H} \partial_i \partial_0 \Phi^{(1)} \partial^i
  \Phi^{(1)} - 4 \partial_i \partial_0\Phi^{(1)} \partial^i \partial_0\Phi^{(1)}) .
\end{eqnarray}\vspace{5mm}\ruledown
\begin{multicols}{2}\hspace{-6.5mm}
Using Eqs.~(\ref{91-0})--(\ref{95}) and substituting the expression of
$\rho^{(\rm{GI}, 2)}$ into the spatial part of the gauge invariant second order
Einstein field equations, we can rewrite the second order Einstein field
equations in Eq.~(\ref{20c}) to be
\begin{equation}
  \mathcal{G}_{i  j} +\mathcal{S}_{i  j} = 0 \label{118c},
\end{equation}
where the tensor $\mathcal{G}_{i  j}$ consists of the
second order metric perturbations while the tensor $\mathcal{S}_{i  j}$
consists of the square of the first order metric perturbations. Here, explicit expressions of them are presented as follows
\end{multicols}
\hspace{0mm}\vspace{5mm}\ruleup
\begin{eqnarray}
  \mathcal{S}_{i  j} & = & \delta_{i  j} \Big( \frac{4
  (c_s^{(2)})^2}{3 (1 + w)} \partial_k \Phi^{(1)} \partial^k \Phi^{(1)} - 12
  (c_s^2 + (c_s^{(2)})^2 - 2w) \mathcal{H}^2 (\Phi^{(1)})^2 - 4 (5 + 3 c_s^2)
  \mathcal{H} \Phi^{(1)} \partial_0 \Phi^{(1)}   \nonumber\\
  &  & + \frac{8 (c_s^{(2)})^2}{3 (1 + w) \mathcal{H}} \partial_k \partial_0
  \Phi^{(1)} \partial^k \Phi^{(1)} - (1 + 3 (c_s^{(2)})^2) (\partial_0
  \Phi^{(1)})^2 - 4 \Phi^{(1)} \partial_0^2 \Phi^{(1)} - 4 \Phi^{(1)} \Delta
  \Phi^{(1)} - 3 \partial_k \Phi^{(1)} \partial^k \Phi^{(1)} \nonumber\\
  &  & - 8 (c_s^{(2)})^2 \Phi^{(1)}\Delta
  \Phi^{(1)}  - 3 (c_s^{(2)})^2 \partial_k \Phi^{(1)} \partial^k \Phi^{(1)} +
  \frac{4 (c_s^{(2)})^2}{3 (1 + w) \mathcal{H}^2} \partial_k \partial_0
  \Phi^{(1)} \partial^k \partial_0 \Phi^{(1)} + 4 c_s^2 \Phi^{(1)} \Delta
  \Phi^{(1)} \Big) \nonumber\\
  &  & + 4 \Phi^{(1)} \partial_i \partial_j \Phi^{(1)} - \frac{4}{3 (1 + w)
  \mathcal{H}} (\partial_i \partial_0 \Phi^{(1)} \partial_j \Phi^{(1)} +
  \partial_i \Phi^{(1)} \partial_0 \partial_j \Phi^{(1)}) \nonumber\\
  &  & + \big( 2 - \frac{4}{3 (1 + w)} \big) \partial_i \Phi^{(1)}
  \partial_j \Phi^{(1)} - \frac{4}{3 (1 + w) \mathcal{H}^2} \partial_i
  \partial_0 \Phi^{(1)} \partial_i \partial_0 \Phi^{(1)},  \label{119c}\\
  \mathcal{G}_{i  j} & = & \frac{1}{4} \partial_0^2 H_{i 
  j}^{(2)} + \frac{1}{2} \mathcal{H} \partial_0 H_{i  j}^{(2)} -
  \frac{1}{4} \Delta H_{i  j}^{(2)} \nonumber\\
  &  & + \delta_{i  j} \big( 3 ((c_s^{(2)})^2 - w) \mathcal{H}^2
  \Phi^{(2)} +\mathcal{H} \partial_0 \Phi^{(2)} + \frac{1}{2} \Delta
  \Phi^{(2)}  \nonumber\\
  &  &  + \partial_0^2 \Psi^{(2)} + (2 + 3 (c_s^{(2)})^2) \mathcal{H}
  \partial_0 \Psi^{(2)} - \frac{1}{2} (1 + 2 (c_s^{(2)})^2) \Delta \Psi^{(2)}
  \big) \nonumber\\
  &  & - \frac{1}{2} \mathcal{H} \partial_i V_j^{(2)} - \frac{1}{4}
  \partial_i \partial_0 V_j^{(2)} - \frac{1}{2} \mathcal{H} \partial_j
  V_i^{(2)} - \frac{1}{4} \partial_j \partial_0 V_i^{(2)} - \frac{1}{2}
  \partial_i \partial_j \Phi^{(2)} + \frac{1}{2} \partial_i \partial_j
  \Psi^{(2)} .  \label{120c}
\end{eqnarray}
We can decompose Eq.~(\ref{118c}) into the tensor, vector and scalar components. 
For illustrations, we decompose $\mathcal{G}_{i  j}$ as a first step. 
The decomposition of $\mathcal{G}_{i  j}$ is explicitly given by
\begin{subequations}
\begin{eqnarray}
  \Lambda_{k  l}^{i  j} \mathcal{G}_{i  j} & = &
  \frac{1}{4} \partial_0^2 H_{k l}^{(2)} + \frac{1}{2} \mathcal{H}
  \partial_0 H_{k l}^{(2)} - \frac{1}{4} \Delta H_{k l}^{(2)} ,\label{124d}\\
  \Delta^{- 1} \partial^l \mathcal{T}^k_i \mathcal{G}_{k  l} & = &
  \frac{1}{4} (\partial_0 + 2\mathcal{H}) V_i^{(2)} ,\\
  \frac{1}{2} \Delta^{- 1} ( \partial^k \Delta^{- 1} \partial^l -
  \frac{1}{2} \mathcal{T}^{k  l} ) \mathcal{G}_{k  l} &
  = & \frac{1}{4} (\Psi^{(2)} - \Phi^{(2)}), \\
  \frac{1}{4} \mathcal{T}^{i  j} \mathcal{G}_{i  j} & = &
  \frac{1}{2} \big( 3 ((c_s^{(2)})^2 - w) \mathcal{H}^2 \Phi^{(2)}
  +\mathcal{H} \partial_0 \Phi^{(2)} + \frac{1}{2} \Delta \Phi^{(2)} 
  \nonumber\\
  &  &  + \partial_0^2 \Psi^{(2)} + (2 + 3 (c_s^{(2)})^2) \mathcal{H}
  \partial_0 \Psi^{(2)} - \frac{1}{2} (1 + 2 (c_s^{(2)})^2) \Delta \Psi^{(2)}
  \big) , \label{127d}
\end{eqnarray}
\end{subequations}
where $\Lambda^{kl}_{ij} \equiv \mathcal{T}^k_i \mathcal{T}^l_j - ({1}/{2})
\mathcal{T}_{ij} \mathcal{T}^{kl}$. The equations of motion of the
gauge invariant second order cosmological perturbations can be written as
\begin{subequations}
\begin{eqnarray}
  \partial_0^2 H_{ij}^{(2)} + 2\mathcal{H} \partial_0 H_{ij}^{(2)} - \Delta
  H_{ij}^{(2)} & = & - 4 \Lambda^{kl}_{ij} \mathcal{S}_{kl},  \label{125c}\\
  (\partial_0 + 2\mathcal{H}) V_i^{(2)} & = & - 4 \Delta^{- 1} \mathcal{T}^k_i
  \partial^l \mathcal{S}_{k  l}  \label{124c},\\
  \Psi^{(2)} - \Phi^{(2)} & = & - 2 \Delta^{- 1} ( \partial^i \Delta^{-
  1} \partial^j - \frac{1}{2} \mathcal{T}^{i  j} )
  \mathcal{S}_{i  j} ,\\
  \partial_0^2 \Psi^{(2)} + (2 + 3 (c_s^{(2)})^2) \mathcal{H} \partial_0
\Psi^{(2)} - \frac{1}{2} (1 + 2 (c_s^{(2)})^2) \Delta \Psi^{(2)}  &  &  \nonumber\\
 +  3 ((c_s^{(2)})^2 - w) \mathcal{H}^2 \Phi^{(2)} +\mathcal{H} \partial_0
  \Phi^{(2)} + \frac{1}{2} \Delta \Phi^{(2)} & = &  -
  \frac{1}{2} \mathcal{T}^{i  j} \mathcal{S}_{i  j} ,
  \label{128c}
\end{eqnarray}
\end{subequations}
where we have the following expressions of $\mathcal{S}_{i j}$, 
\begin{eqnarray}
  \Lambda^{i  j}_{k  l} \mathcal{S}_{i  j} & = &
  \Lambda^{i  j}_{k  l} \Big( \big( 2 + \frac{4}{3 (1 + w)}
  \big) \Phi^{(1)} \partial_i \partial_j \Phi^{(1)} + \frac{8}{3 (1 + w)
  \mathcal{H}} \Phi^{(1)} \partial_0 \partial_i \partial_j \Phi^{(1)} \nonumber\\
&  &-
  \frac{4}{3 (1 + w) \mathcal{H}^2} \partial_i \partial_0 \Phi^{(1)}
  \partial_j \partial_0 \Phi^{(1)} \Big)  \label{129c}, \\
  \Delta^{- 1} \mathcal{T}^k_i \partial^l \mathcal{S}_{k  l} & = &
  \Delta^{- 1} \mathcal{T}^k_i \partial^l \Big( ( 2 - \frac{4}{3 (1 +
  w)} ) \partial_k \Phi^{(1)} \partial_l \Phi^{(1)} - \frac{4}{3 (1 + w) \mathcal{H}} (\partial_k \partial_0
\Phi^{(1)} \partial_l \Phi^{(1)} + \partial_k \Phi^{(1)} \partial_0
\partial_l \Phi^{(1)})  \nonumber\\
  &  & + 4 \Phi^{(1)}
  \partial_k \partial_l \Phi^{(1)}   - \frac{4}{3 (1 + w) \mathcal{H}^2} \partial_k
  \partial_0 \Phi^{(1)} \partial_l \partial_0 \Phi^{(1)} \Big) ,\\
  \frac{1}{4} \mathcal{T}^{i  j} \mathcal{S}_{i  j} & = &
  \frac{1}{2} \Bigg( \frac{4 (c_s^{(2)})^2}{3 (1 + w)} \partial_k \Phi^{(1)}
  \partial^k \Phi^{(1)} - 12 (c_s^2 + (c_s^{(2)})^2 - 2w) \mathcal{H}^2
  (\Phi^{(1)})^2 - 4 (5 + 3 c_s^2) \mathcal{H} \Phi^{(1)} \partial_0
  \Phi^{(1)}  \nonumber\\
  &  & - 8 (c_s^{(2)})^2 \Phi^{(1)}\Delta \Phi^{(1)}  + \frac{8 (c_s^{(2)})^2}{3 (1 + w) \mathcal{H}} \partial_k \partial_0
  \Phi^{(1)} \partial^k \Phi^{(1)} - (1 + 3 (c_s^{(2)})^2) (\partial_0
  \Phi^{(1)})^2 - 4 \Phi^{(1)} \partial_0 ^2\Phi^{(1)}  \nonumber\\
  &  &  - 4 \Phi^{(1)} \Delta
  \Phi^{(1)} - 3 \partial_k \Phi^{(1)} \partial^k \Phi^{(1)} - 3 (c_s^{(2)})^2 \partial_k \Phi^{(1)} \partial^k \Phi^{(1)} +
  \frac{4 (c_s^{(2)})^2}{3 (1 + w) \mathcal{H}^2} \partial_k \partial_0
  \Phi^{(1)} \partial^k \partial_0 \Phi^{(1)}  \nonumber\\
  &  & + 4 c_s^2 \Phi^{(1)} \Delta
  \Phi^{(1)} \Bigg) + \frac{1}{4} \mathcal{T}^{i  j} \Bigg( \big( 2 + \frac{4}{3 (1
  + w)} \big) \Phi^{(1)} \partial_i \partial_j \Phi^{(1)} + \frac{8}{3 (1 +
  w) \mathcal{H}} \Phi^{(1)} \partial_0 \partial_i \partial_j \Phi^{(1)} \nonumber\\
& & -  \frac{4}{3 (1 + w) \mathcal{H}^2} \partial_i \partial_0 \Phi^{(1)}
  \partial_j \partial_0 \Phi^{(1)} \Bigg) ,
\end{eqnarray}
\vspace{-0.3cm}
\begin{eqnarray}
  \frac{1}{2} \Delta^{- 1} \big( \partial^i \Delta^{- 1} \partial^j -
  \frac{1}{2} \mathcal{T}^{i  j} \big) \mathcal{S}_{i  j} &
  = & \frac{1}{2} \Delta^{- 1} \big( \partial^i \Delta^{- 1} \partial^j -
  \frac{1}{2} \mathcal{T}^{i  j} \big) \left( \big( 2 - \frac{4}{3
  (1 + w)} \big) \partial_i \Phi^{(1)} \partial_j \Phi^{(1)} + 4 \Phi^{(1)}
  \partial_i \partial_j \Phi^{(1)} \right. \nonumber\\
  &  & - \frac{4}{3 (1 + w) \mathcal{H}} (\partial_i \partial_0 \Phi^{(1)}
  \partial_j \Phi^{(1)} + \partial_i \Phi^{(1)} \partial_0 \partial_j
  \Phi^{(1)}) \nonumber\\
  &  & \left. - \frac{4}{3 (1 + w) \mathcal{H}^2} \partial_i \partial_0
  \Phi^{(1)} \partial_j \partial_0 \Phi^{(1)} \right) . \label{132c}
\end{eqnarray}\vspace{5mm}\ruledown
\begin{multicols}{2}\hspace{-6.5mm}
The above equations of motion can be derived in an easier way, by adopting the scalar-vector-tensor decomposition operators onto both sides of Eq.~(\ref{118c}). 
These operators are defined on the background and thus gauge invariant. 
Therefore, the equations of motion in Eqs.~(\ref{125c})--(\ref{128c}) should also be gauge invariant. 
Here, the operator $\Lambda^{kl}_{ij}$ is defined in
configuration space. In Appendix~\ref{F}, we  express $\Lambda^{kl}_{ij}$
in momentum space and show that it can be rewritten in terms of the polarization tensors. 
%From Eqs.~(\ref{125c})--(\ref{128c}), the second order cosmological perturbations (tensor perturbations)evolve independently of the other types of the second order metric perturbations (scalar and vector perturbations), however, they all could be induced by first-order scalar perturbations. This mightsuggest that scalar, vector and tensor perturbations does not entirely getentangled with the perturbations of other types, due to the Einstein fieldequations.

\subsubsection{Gauge invariant second order synchronous variables}

At second order, the gauge invariant synchronous  metric perturbations  are given by
\begin{eqnarray}
  g_{\mu \nu}^{(\mathrm{\rm{GI}, 2)}} \mathrm{d} x^{\mu} \mathrm{d} x^{\nu}
  & = & a^2  (- 2 \delta_{ij} \Psi^{(2)} + 2 \partial_i \partial_j
  E^{(2)} \nonumber\\
  & & + \partial_i C_j^{(2)} + \partial_j
  C_i^{(2)} + H_{ij}^{(2)}) \mathrm{d} x^i \mathrm{d} x^j . 
\end{eqnarray}
The explicit expressions of the gauge invariant second order  variables,
$E^{(2)}$, $\Psi^{(2)}$, $C_i^{(2)}$ and $H_{ij}^{(2)}$ are
presented in Eqs.~(\ref{84c})--(\ref{90}) with the expression of
$\mathcal{X}_{\mu \nu}^N$ in Appendix \ref{D}. Based on the temporal components of the
second order Einstein field equations and Eqs.~(\ref{91-0})--(\ref{95}),
the gauge invariant second order density perturbations in terms of the metric
perturbations is given by
\end{multicols}
\hspace{0mm}\vspace{5mm}\ruleup
\begin{eqnarray}
  \rho^{(\rm{GI}, 2)} & = & \frac{2}{3 (1 + w) \kappa a^2 \mathcal{H}^2} (36
  (1 + w) \mathcal{H}^4 (\Phi^{(1)})^2 - 3 (1 + w) \mathcal{H}^3 (3 \partial_0
  \Psi^{(2)} - \Delta \partial_0 E^{(2)})  \nonumber\\
  &  & +\mathcal{H}^2 (9 (1 + w) (\partial_0 \Phi^{(1)})^2 + 24 (1 + w)
  \Phi^{(1)} \Delta \Phi^{(1)} + 3 (1 + w) \Delta \Psi^{(2)} + (5 + 9 w)
  \partial_i \Phi^{(1)} \partial^i \Phi^{(1)}) \nonumber\\
  &  &  - 8\mathcal{H} \partial_i \partial_0 \Phi^{(1)} \partial^i
  \Phi^{(1)} - 4 \partial_i \partial_0\Phi^{(1)} \partial^i \partial_0\Phi^{(1)}) .
\end{eqnarray}\vspace{5mm}\ruledown
\begin{multicols}{2}\hspace{-6.5mm}
Using Eqs.~(\ref{91-0})--(\ref{95}) and substituting the expression of
$\rho^{(\rm{GI}, 2)}$ into the spatial parts of the gauge invariant second order 
Einstein field equations, we can also rewrite the second order Einstein field
equations in the terms of the tensors $\mathcal{G}_{i  j}$ and
$\mathcal{S}_{i  j}$, i.e.,
\begin{equation}
  \mathcal{G}_{i  j} +\mathcal{S}_{i  j} = 0,\label{eq:saisyn1}
\end{equation}
where the $\mathcal{S}_{i  j}$ takes the same form as Eq.~(\ref{119c}) and
\end{multicols}
\hspace{0mm}\vspace{5mm}\ruleup
\begin{eqnarray}
  \mathcal{G}_{i  j} & = & \frac{1}{4} \partial_0^2 H_{i 
  j}^{(2)} + \frac{1}{2} \mathcal{H} \partial_0 H_{i  j}^{(2)} -
  \frac{1}{4} \Delta H_{i  j}^{(2)} \nonumber\\
  &  & + \delta_{i  j} \Big( \partial_0^2 \Psi^{(2)} + (2 + 3
  (c_s^{(2)})^2) \mathcal{H} \partial_0 \Psi^{(2)}  - \frac{1}{2} (1 + 2
  (c_s^{(2)})^2) \Delta \Psi^{(2)}    - \Delta \big( \frac{1}{2} (\partial_0 + 2\mathcal{H})
  \partial_{\eta} E^{(2)} + (c_s^{(2)})^2 \mathcal{H} \partial_{\eta}
  E^{(2)} \big) \Big) \nonumber\\
  &  & + \frac{1}{2} \mathcal{H} \partial_i \partial_0 C_j^{(2)} +
  \frac{1}{4} \partial_i \partial_0^2 C_j^{(2)} + \frac{1}{2}
  \mathcal{H} \partial_j \partial_0 C_i^{(2)}  + \frac{1}{4}
  \partial_j \partial_0^2 C_i^{(2)} +\mathcal{H} \partial_i
  \partial_j \partial_0 E^{(2)} + \frac{1}{2} \partial_j \partial_j
  \partial_0^2 E^{(2)} + \frac{1}{2} \partial_i \partial_j \Psi^{(2)}
  .  \label{136c}
\end{eqnarray}
By adopting the scalar-vector-tensor decomposition to Eq.~(\ref{eq:saisyn1}), the equations of motion of the gauge invariant second order cosmological perturbations are obtained to be

\begin{subequations}
\begin{eqnarray}
  \partial_0^2 H_{i j}^{(2)} + 2\mathcal{H} \partial_0 H_{ i j}^{(2)} - \Delta H_{i j}^{(2)} & = & - 4
  \Lambda^{kl}_{ij} \mathcal{S}_{kl},  \label{137a}\\
  (\partial_0 + 2\mathcal{H}) \partial_0 C_i^{(2)} & = & - 4
  \Delta^{- 1} \mathcal{T}^k_i \partial^l \mathcal{S}_{k  l}, \label{137b}\\
  (\partial_0 + 2\mathcal{H}) \partial_0 E^{(2)} + \Psi^{(2)} & = & -
  2 \Delta^{- 1} \big( \partial^i \Delta^{- 1} \partial^j - \frac{1}{2}
  \mathcal{T}^{i  j} \big) \mathcal{S}_{i  j},  \label{137c} \\
  \partial_0^2 \Psi^{(2)} + (2 + 3 (c_s^{(2)})^2) \mathcal{H} \partial_0
  \Psi^{(2)} - \frac{1}{2} (1 + 2 (c_s^{(2)})^2) \Delta \Psi^{(2)} &  & 
  \nonumber\\
  - \Delta \big( \frac{1}{2} (\partial_0 + 2\mathcal{H}) \partial_0
  E^{(2)} + (c_s^{(2)})^2 \mathcal{H} \partial_0 E^{(2)}
  \big) & = & - \frac{1}{2} \mathcal{T}^{i  j} \mathcal{S}_{i
   j} . 
\end{eqnarray}
\end{subequations}
\vspace{5mm}\ruledown
\begin{multicols}{2}\hspace{-6.5mm}
The right hand sides of the above equations are as same as those in Eqs.~(\ref{129c})--(\ref{132c}), since the source term $\mathcal{S}_{i  j}$ is uniquely determined by the gauge invariant first order Newtonian variables. 
In particular, the equation of motion of the tensor perturbation $H_{\mu \nu}^{(\rm{GI},2)}$ in Eq.~(\ref{137a}) is as same as that in Eq.~(\ref{125c}). 
For the vector perturbation, the evolution of $\partial_0 C_i^{(2)}$ in Eq.~(\ref{137b}) is shown to be
as same as that of $V_i^{(2)}$ in Eq.~(\ref{124c}).

\subsection{Second order cosmological perturbations induced by the first order synchronous variables}

In the subsection, we turn to discuss the gauge invariant synchronous metric perturbations at first order
and their equations of motion.  For the second order equations of motion, we will also consider the gauge invariant variables that are Newtonian and
synchronous, respectively. 

The gauge invariant synchronous  metric up to first order is given by
\begin{eqnarray}
  g_{\mu \nu}^{(\mathrm{\rm{GI})}} \mathrm{d} x^{\mu} \mathrm{d} x^{\nu} & =
  & - a^2 \mathrm{d} \eta^2 + a^2  (\delta_{ij} (1 - 2 \Psi^{(1)}) + 2
  \partial_i \partial_j E^{(1)}   \nonumber\\
  & & + \partial_i
  C_j^{(1)} + \partial_j C_i^{(1)}) \mathrm{d} x^i
  \mathrm{d} x^j ,  \label{99}
\end{eqnarray}
where we also neglect the first order tensor perturbation, i.e., $H_{i j}^{(1)}=0$. Here, the gauge invariant first order variables $\Psi^{(1)}$, $E^{(1)}$ and
$C^{(1)}_i$ have been shown in Eqs.~(\ref{44})--(\ref{47}). Based on
the first-order Einstein field equations and Eqs.~(\ref{91-0}) and (\ref{91}), the density and velocity perturbations are expressed in terms of the gauge invariant first order metric perturbations, i.e.,
\begin{subequations}
\begin{eqnarray}
  \rho^{(\mathrm{\rm{GI}, 1)}} & = & \frac{- 6\mathcal{H} \partial_0
  \Psi^{(1)} + 2\mathcal{H} \Delta \partial_0 E^{(1)} + 2 \Delta
  \Psi^{(1)}}{\kappa a^2},  \label{101}\\
  \upsilon_i^{(\mathrm{\rm{GI}, 1)}} & = & - \frac{\Delta \partial_0
  C_i^{(1)} + 4 \partial_i \partial_0 \Psi^{(1)}}{6 (1 + w)
  \mathcal{H}^2} . 
\end{eqnarray}
\end{subequations}
By substituting Eqs.~(\ref{91-0}) and (\ref{91}) and the equations above into the spatial parts of the first-order Einstein field equations, we can obtain 
\begin{eqnarray}
  C_i^{(1)} & = & 0,  \label{144c}
\end{eqnarray}
and the equations of motion of $E^{(1)}$ and $\Psi^{(1)}$, i.e.,
\begin{subequations}
\begin{eqnarray}
  \Psi^{(1)} + 2\mathcal{H} \partial_0 E^{(1)} + \partial_0^2
  E^{(1)} & = & 0,  \label{104}\\
  \partial_0^2 \Psi^{(1)} +\mathcal{H} (2 + 3 c_s^2) \partial_0 \Psi^{(1)} & &\nonumber\\
   -
  c_s^2 \Delta \Psi^{(1)} -\mathcal{H}c_s^2 \Delta \partial_0 E^{(1)}
  & = & 0. 
\end{eqnarray}
\end{subequations}
Here, we also disregard the decaying mode of the first order vector perturbations.

\subsubsection{Gauge invariant second order Newtonian  variables}

At second order, the gauge invariant Newtonian  metric perturbations are given by
\begin{eqnarray}
  g_{\mu \nu}^{(\mathrm{\rm{GI}, 2)}} \mathrm{d} x^{\mu} \mathrm{d} x^{\nu}
  & = & 2 a^2 \Phi^{(2)} \mathrm{d} \eta^2 + 2 a^2 V_i^{(2)} \mathrm{d} \eta
  \mathrm{d} x^i \nonumber\\
  & & + a^2  (- 2 \delta_{ij} \Psi^{(2)} + H_{ij}^{(2)}) \mathrm{d}
  x^i \mathrm{d} x^j,  
\end{eqnarray}
The explicit expressions of the gauge invariant second order variables,
$\Phi^{(2)}$, $\Psi^{(2)}$, $V_i^{(2)}$ and $H_{ij}^{(2)}$ are presented in
Eqs.~(\ref{74c})--(\ref{67}) with the expression of $\mathcal{X}_{\mu \nu}^S$
in Appendix \ref{D}. Based on the temporal components of the second order Einstein
field equations and Eqs.~(\ref{91-0}), (\ref{91}),
(\ref{101})--(\ref{144c}), we can express the gauge invariant second order 
 density perturbation in terms of metric perturbations, namely,
\end{multicols}
\hspace{0mm}\vspace{5mm}\ruleup
\begin{eqnarray}
  \rho^{(\rm{GI}, 2)} & = & - \frac{1}{3 (1 + w) \kappa a^2 \mathcal{H}^2}
  \Big(18 (1 + w) \mathcal{H}^4 \Phi^{(2)} + 8 \partial_i \partial_0 \Psi^{(1)}
  \partial^i \partial_0 \Psi^{(1)} \nonumber\\
  &  & + 6 (1 + w) \mathcal{H}^3 (3 \partial_0 \Psi^{(2)} - 4 \partial_0
  \Psi^{(1)} \Delta E^{(1)} + 4 \Psi^{(1)} (3 \partial_0 \Psi^{(1)} -
  \Delta \partial_0 {e}^{(1)}) + 4 \partial_i \partial_j \partial_0
  E^{(1)} \partial^i \partial^j E^{(1)})\nonumber\\
  &  & - 3 (1 + w) \mathcal{H}^2 (6 (\partial_0 \Psi^{(1)})^2 - 4 \partial_0
  \Psi^{(1)} \Delta \partial_0 E^{(1)} + 16 \Psi^{(1)} \Delta
  \Psi^{(1)} + 2 \Delta \Psi^{(2)} + 6 \partial_i \Psi^{(1)} \partial^i
  \Psi^{(1)} 
\nonumber\\
  &  & - 4 \partial_i \Delta E^{(1)} \partial^i \Psi^{(1)} + \Delta \partial_0 E^{(1)} \Delta \partial_0
  E^{(1)} - 4 \Delta E^{(1)} \Delta \Psi^{(1)} - \partial_i
  \Delta E^{(1)} \partial^i \Delta E^{(1)} \nonumber\\
  &  &   - 4 \partial_i
  \partial_j \Psi^{(1)} \partial^i \partial^j E^{(1)}  - \partial_i \partial_j \partial_0
  E^{(1)} \partial^i \partial^j \partial_0 E^{(1)} +
  \partial_k \partial_i \partial_j E^{(1)} \partial^i \partial^j
  \partial^k E^{(1)})\Big).
\end{eqnarray}
\vspace{5mm}\ruledown
\begin{multicols}{2}\hspace{-6.5mm}
Using Eqs.~(\ref{91-0}), (\ref{91}),(\ref{101})--(\ref{144c}) and
substituting the expression of $\rho^{(\rm{GI}, 2)}$ into the spatial parts of the
gauge invariant second order  Einstein field equations, we can
also obtain 
\begin{equation}
  \mathcal{G}_{i  j} +\mathcal{S}_{i  j} = 0,
\end{equation}
where the form of $\mathcal{G}_{i  j}$ is the same as Eq.~(\ref{120c})
and
\end{multicols}
\hspace{0mm}\vspace{5mm}\ruleup
\begin{eqnarray}
  \mathcal{S}_{i  j} & = & \delta_{i  j} \Big( - 12 (c_s^2 -
  (c_s^{(2)})^2) \mathcal{H} \Psi^{(1)} \partial_0 \Psi^{(1)} + ( 1 - 3
  ( c_s^{(2)} )^2 ) (\partial_0 \Psi^{(1)})^2 - 2 (1 - 2
  c_s^2 + 4 (c_s^{(2)})^2) \Psi^{(1)} \Delta \Psi^{(1)} \nonumber\\
  &  & - (2 + 3 (c_s^{(2)})^2) \partial_k \Psi^{(1)} \partial^k \Psi^{(1)} +
  \frac{4 (c_s^{(2)})^2}{3 (1 + w) \mathcal{H}^2} \partial_k \partial_0
  \Psi^{(1)} \partial^k \partial_0 \Psi^{(1)} - (1 + (c_s^{(2)})^2) \Delta
  \partial_0 E^{(1)} \Delta \partial_0 E^{(1)}\nonumber\\
  &  & + (1 + (c_s^{(2)})^2) \partial_k \Delta E^{(1)} \partial^k
  \Delta E^{(1)} + 8 (1 + (c_s^{(2)})^2) \mathcal{H} \partial_k
  \partial_l \partial_0 E^{(1)} \partial^k \partial^l
  E^{(1)} + 4 \partial_k \partial_l \partial_0^2 E^{(1)}
  \partial^k \partial^l E^{(1)}\nonumber\\
  &  & + (3 + (c_s^{(2)})^2) \partial_k \partial_l \partial_0
  E^{(1)} \partial^k \partial^l \partial_0 E^{(1)} - (1 +
  (c_s^{(2)})^2) \partial_m \partial_l \partial_k E^{(1)} \partial^m
  \partial^l \partial^k E^{(1)} \nonumber\\
  &  & - 2 \partial_0^2 \Psi^{(1)} \Delta E^{(1)} - (1 - 2
  (c_s^{(2)})^2) \partial_0 \Psi^{(1)} \Delta \partial_0 E^{(1)} + (1
  + 2 (c_s^{(2)})^2) \partial_k \Delta E^{(1)} \partial^k \Psi^{(1)}
 \nonumber\\
  &  &   + 2 (1 + (c_s^{(2)})^2) \Delta E^{(1)} \Delta \Psi^{(1)} + 2 (c_s^{(2)})^2 \partial_k \partial_l \Psi^{(1)} \partial^k
  \partial^l E^{(1)} \nonumber\\
  &  & - 2 - 4\mathcal{H} \big( \partial_0 \Psi^{(1)}
  ( 1 + (c_s^{(2)})^2 ) \Delta E^{(1)} + \Psi^{(1)}
  ((c_s^{(2)})^2 - c_s^2) \Delta \partial_0 E^{(1)} \big) \Big)\nonumber\\
  &  & + 3 \partial_i \Psi^{(1)} \partial_j \Psi^{(1)} - \frac{4}{3 (1 + w)
  \mathcal{H}^2} \partial_i \partial_0 \Psi^{(1)} \partial_j \partial_0
  \Psi^{(1)} + 2 \Psi^{(1)} \partial_i \partial_j \Psi^{(1)}\nonumber\\
  &  & - 2 \partial_k \partial_j \partial_0 E^{(1)} \partial^k
  \partial_i \partial_0 E^{(1)} - \partial_k \Delta E^{(1)}
  \partial^k \partial_i \partial_j E^{(1)} + \partial_k \partial_l
  \partial_j E^{(1)} \partial^k \partial^l \partial_i
  E^{(1)}\nonumber\\
  &  & - 2 \Delta \partial_0^2 E^{(1)} \partial_i \partial_j
  E^{(1)} - 4\mathcal{H} (1 + c_s^2) \Delta \partial_0
  E^{(1)} \partial_i \partial_j E^{(1)} + \Delta \partial_0
  E^{(1)} \partial_i \partial_j \partial_0 E^{(1)}\nonumber\\
  &  & - \partial_k \partial_i \partial_j E^{(1)} \partial^k
  \Psi^{(1)} + 2 \partial_k \partial_j \Psi^{(1)} \partial^k \partial_i
  E^{(1)} + 2 \partial_k \partial_i \Psi^{(1)} \partial^k \partial_j
  E^{(1)} + 12 (1 + c_s^2) \mathcal{H} \partial_0 \Psi^{(1)}
  \partial_i \partial_j E^{(1)}\nonumber\nonumber\\
  &  & + 6 \partial_0^2 \Psi^{(1)} \partial_i \partial_j E^{(1)} - 4
  (1 + c_s^2) \Delta \Psi^{(1)} \partial_i \partial_j E^{(1)} +
  \partial_0 \Psi^{(1)} \partial_i \partial_j \partial_0 E^{(1)} - 2
  \Delta E^{(1)} \partial_i \partial_j \Psi^{(1)} \label{149c}.
\end{eqnarray}
Following the approach in the previous subsection, the equations of motion of
the gauge invariant second order cosmological perturbations can be written as
\begin{subequations}
\begin{eqnarray}
  \partial_0^2 H_{ij}^{(2)} + 2\mathcal{H} \partial_0 H_{ij}^{(2)} - \Delta
  H_{ij}^{(2)} & = & - 4 \Lambda^{kl}_{ij} \mathcal{S}_{kl}, \label{eq:sai147}\\
  (\partial_0 + 2\mathcal{H}) V_i^{(2)} & = & - 4 \Delta^{- 1} \mathcal{T}^k_i
  \partial^l \mathcal{S}_{k  l} ,\label{eq:sai148}\\
  \Psi^{(2)} - \Phi^{(2)} & = & - 2 \Delta^{- 1} \big( \partial^i \Delta^{-
  1} \partial^j - \frac{1}{2} \mathcal{T}^{i  j} \big)
  \mathcal{S}_{i  j} ,\\
   \partial_0^2 \Psi^{(2)} + (2 + 3 (c_s^{(2)})^2) \mathcal{H} \partial_0
\Psi^{(2)} - \frac{1}{2} (1 + 2 (c_s^{(2)})^2) \Delta \Psi^{(2)} 
&  &  \nonumber\\
 + 3 ((c_s^{(2)})^2 - w) \mathcal{H}^2 \Phi^{(2)} +\mathcal{H} \partial_0
\Phi^{(2)} + \frac{1}{2} \Delta \Phi^{(2)} & = & -
  \frac{1}{2} \mathcal{T}^{i  j} \mathcal{S}_{i  j} ,
\end{eqnarray}
\end{subequations}
where
\begin{eqnarray}
  \Lambda^{i  j}_{k  l} \mathcal{S}_{i  j} & = &
  \Lambda^{i  j}_{k  l} \Big( -\Psi^{(1)} \partial_i \partial_j \Psi^{(1)} +
  \frac{4}{3 (1 + w) \mathcal{H}^2} \partial_0 \Psi^{(1)} \partial_i
  \partial_j \partial_0 \Psi^{(1)}\nonumber\\
  &  & + (- 5 \partial^m \Psi^{(1)} \partial_m - 6 c_s^2 \mathcal{H} \partial_0
  \Psi^{(1)} - 2 (1 - c_s^2) \Delta \Psi^{(1)} + \partial_0 \Psi^{(1)}
  \partial_0 - 2 \Psi^{(1)} \Delta) \partial_i \partial_j E^{(1)}\nonumber\\
  &  & + 2 \partial_m \partial_0 E^{(1)} \partial^m \partial_i
  \partial_j \partial_0 E^{(1)} + \Delta \partial_0 E^{(1)}
  \partial_i \partial_j \partial_0 E^{(1)} - \partial_m \Delta
  E^{(1)} \partial^m \partial_i \partial_j E^{(1)}\nonumber\\
  &  & - \partial_n \partial_m E^{(1)} \partial^n \partial^m
  \partial_i \partial_j E^{(1)} + 2 c_s^2 \mathcal{H} \Delta
  \partial_0 E^{(1)} \partial_i \partial_j E^{(1)} \Big) ,\label{eq:sai151}\\
  \Delta^{- 1} \mathcal{T}^i_m \partial^j \mathcal{S}_{i  j} & = &
  \Delta^{- 1} \mathcal{T}^i_m \partial^j \Big( 3 \partial_i \Psi^{(1)}
  \partial_j \Psi^{(1)} - \frac{4}{3 (1 + w) \mathcal{H}^2} \partial_i
  \partial_0 \Psi^{(1)} \partial_j \partial_0 \Psi^{(1)} + 2 \Psi^{(1)}
  \partial_i \partial_j \Psi^{(1)}  \nonumber\\
  &  & - 2 \partial_k \partial_j \partial_0 E^{(1)} \partial^k
  \partial_i \partial_0 E^{(1)} - \partial_k \Delta E^{(1)}
  \partial^k \partial_i \partial_j E^{(1)} + \partial_k \partial_l
  \partial_j E^{(1)} \partial^k \partial^l \partial_i
  E^{(1)} \nonumber\\
  &  & - 2 \Delta \partial_0^2 E^{(1)} \partial_i \partial_j
  E^{(1)} - 4\mathcal{H} (1 + c_s^2) \Delta \partial_0
  E^{(1)} \partial_i \partial_j E^{(1)} + \Delta \partial_0
  E^{(1)} \partial_i \partial_j \partial_0 E^{(1)}
  \nonumber\\
  &  & - \partial_k \partial_i \partial_j E^{(1)} \partial^k
  \Psi^{(1)} + 2 \partial_k \partial_j \Psi^{(1)} \partial^k \partial_i
  E^{(1)} + 2 \partial_k \partial_i \Psi^{(1)} \partial^k \partial_j
  E^{(1)}  \nonumber\\
  &  &  + 12 (1 + c_s^2) \mathcal{H} \partial_0 \Psi^{(1)}
  \partial_i \partial_j E^{(1)} + 6 \partial_0^2 \Psi^{(1)} \partial_i \partial_j
  E^{(1)} - 4 (1 + c_s^2) \Delta \Psi^{(1)} \partial_i \partial_j
  E^{(1)} \nonumber\\
  &  &   + \partial_0 \Psi^{(1)} \partial_i \partial_j \partial_0
  E^{(1)} - 2 \Delta E^{(1)} \partial_i \partial_j
  \Psi^{(1)} \Big) ,\\
  \frac{1}{4} \mathcal{T}^{i  j} \mathcal{S}_{i  j} & = &
  \frac{1}{2} \Big( - 12 (c_s^2 - (c_s^{(2)})^2) \mathcal{H} \Psi^{(1)}
  \partial_0 \Psi^{(1)} + ( 1 - 3 ( c_s^{(2)} )^2 )
  (\partial_0 \Psi^{(1)})^2  \nonumber\\
  &  & - 2 (1 - 2 c_s^2 + 4 (c_s^{(2)})^2) \Psi^{(1)}
  \Delta \Psi^{(1)}  - (2 + 3 (c_s^{(2)})^2) \partial_k \Psi^{(1)} \partial^k \Psi^{(1)} \nonumber\\
  &  & +  \frac{4 (c_s^{(2)})^2}{3 (1 + w) \mathcal{H}^2} \partial_k \partial_0
  \Psi^{(1)} \partial^k \partial_0 \Psi^{(1)} - (1 + (c_s^{(2)})^2) \Delta
  \partial_0 E^{(1)} \Delta \partial_0 E^{(1)}  \nonumber\\
  &  &+ (1 + (c_s^{(2)})^2) \partial_k \Delta E^{(1)} \partial^k
  \Delta E^{(1)} + 8 (1 + (c_s^{(2)})^2) \mathcal{H} \partial_k
  \partial_l \partial_0 E^{(1)} \partial^k \partial^l
  E^{(1)} \nonumber\\
  &  & + 4 \partial_k \partial_l \partial_0^2 E^{(1)}
  \partial^k \partial^l E^{(1)} + (3 + (c_s^{(2)})^2) \partial_k \partial_l \partial_0
  E^{(1)} \partial^k \partial^l \partial_0 E^{(1)} \nonumber\\
  &  & - (1 +
  (c_s^{(2)})^2) \partial_m \partial_l \partial_k E^{(1)} \partial^m
  \partial^l \partial^k E^{(1)} - 2 \partial_0^2 \Psi^{(1)} \Delta E^{(1)}\nonumber\\
  &  &  - (1 - 2
  (c_s^{(2)})^2) \partial_0 \Psi^{(1)} \Delta \partial_0 E^{(1)} + (1
  + 2 (c_s^{(2)})^2) \partial_k \Delta E^{(1)} \partial^k \Psi^{(1)}
  \nonumber\\
  &  &   + 2 (1 + (c_s^{(2)})^2) \Delta E^{(1)} \Delta \Psi^{(1)}+ 2 (c_s^{(2)})^2 \partial_k \partial_l \Psi^{(1)} \partial^k
  \partial^l E^{(1)} \nonumber\\
  &  &- 4\mathcal{H} \big( \partial_0 \Psi^{(1)}
  ( 1 + (c_s^{(2)})^2 ) \Delta E^{(1)} + \Psi^{(1)}
  ((c_s^{(2)})^2 - c_s^2) \Delta \partial_0 E^{(1)} \big) \Big)
  \nonumber\\
  &  & \frac{1}{4} \mathcal{T}^{i  j} \Big( -\Psi^{(1)} \partial_i \partial_j \Psi^{(1)} +
  \frac{4}{3 (1 + w) \mathcal{H}^2} \partial_0 \Psi^{(1)} \partial_i
  \partial_j \partial_0 \Psi^{(1)}\nonumber\\
  &  & +(- 5 \partial^m \Psi^{(1)} \partial_m - 6 c_s^2 \mathcal{H} \partial_0
  \Psi^{(1)} - 2 (1 - c_s^2) \Delta \Psi^{(1)} + \partial_0 \Psi^{(1)}
  \partial_0 - 2 \Psi^{(1)} \Delta) \partial_i \partial_j E^{(1)}\nonumber\\
  &  & + 2 \partial_m \partial_0 E^{(1)} \partial^m \partial_i
  \partial_j \partial_0 E^{(1)} + \Delta \partial_0 E^{(1)}
  \partial_i \partial_j \partial_0 E^{(1)} - \partial_m \Delta
  E^{(1)} \partial^m \partial_i \partial_j E^{(1)}\nonumber\\
  &  & - \partial_n \partial_m E^{(1)} \partial^n \partial^m
  \partial_i \partial_j E^{(1)} + 2 c_s^2 \mathcal{H} \Delta
  \partial_0 E^{(1)} \partial_i \partial_j E^{(1)}  \Big) ,
\end{eqnarray}
\vspace{-0.3cm}
\begin{eqnarray}
  \frac{1}{2} \Delta^{- 1} \big( \partial^i \Delta^{- 1} \partial^j -
  \frac{1}{2} \mathcal{T}^{i  j} \big) \mathcal{S}_{i  j} &
  = & \frac{1}{2} \Delta^{- 1} \big( \partial^i \Delta^{- 1} \partial^j -
  \frac{1}{2} \mathcal{T}^{i  j} \big) \Big( 3 \partial_i
  \Psi^{(1)} \partial_j \Psi^{(1)} - \frac{4}{3 (1 + w) \mathcal{H}^2}
  \partial_i \partial_0 \Psi^{(1)} \partial_j \partial_0 \Psi^{(1)}   \nonumber\\
  &  &  + 2  \Psi^{(1)} \partial_i \partial_j \Psi^{(1)} - 2 \partial_k \partial_j \partial_0 E^{(1)} \partial^k
  \partial_i \partial_0 E^{(1)} - \partial_k \Delta E^{(1)}
  \partial^k \partial_i \partial_j E^{(1)} \nonumber\\
  &  & + \partial_k \partial_l
  \partial_j E^{(1)} \partial^k \partial^l \partial_i
  E^{(1)}  - 2 \Delta \partial_0^2 E^{(1)} \partial_i \partial_j
  E^{(1)} - 4\mathcal{H} (1 + c_s^2) \Delta \partial_0
  E^{(1)} \partial_i \partial_j E^{(1)} 
  \nonumber\\
  &  & + \Delta \partial_0
  E^{(1)} \partial_i \partial_j \partial_0 E^{(1)} - \partial_k \partial_i \partial_j E^{(1)} \partial^k
  \Psi^{(1)} + 2 \partial_k \partial_j \Psi^{(1)} \partial^k \partial_i
  E^{(1)}  \nonumber\\
  &  &  + 2 \partial_k \partial_i \Psi^{(1)} \partial^k \partial_j
  E^{(1)} + 12 (1 + c_s^2) \mathcal{H} \partial_0 \Psi^{(1)}
  \partial_i \partial_j E^{(1)} + 6 \partial_0^2 \Psi^{(1)} \partial_i \partial_j
  E^{(1)} \nonumber\\
  &  &  - 4 (1 + c_s^2) \Delta \Psi^{(1)} \partial_i \partial_j
  E^{(1)} + \partial_0 \Psi^{(1)} \partial_i \partial_j \partial_0
  E^{(1)} - 2 \Delta E^{(1)} \partial_i \partial_j
  \Psi^{(1)}\Big) .\label{eq:sai154}
\end{eqnarray}
\vspace{5mm}\ruledown
\begin{multicols}{2}\hspace{-6.5mm}
It is not surprising that the left hand sides of the equations of motion are as
same as those of Eqs.~(\ref{125c})--(\ref{128c}). %Using the synchronous gauge invariant variables at the first-order, the scalar-scalar source $\mathcal{S}_{i  j}$ are shown to be more complex.
As suggested in Ref.~\cite{Lu:2020diy}, we show that the source terms in Eq.~(\ref{eq:sai148}) include the  first order perturbation $E^{(1)}$ without temporal derivative. This is different from the formulae in the other previous works \cite{Gong:2019mui,DeLuca:2019ufz,Yuan:2019fwv}.

\subsubsection{Gauge invariant second order synchronous variables}

For the gauge invariant synchronous metric perturbations at second order, we
also have
\begin{eqnarray}
  g_{\mu \nu}^{(\mathrm{\rm{GI}, 2)}} \mathrm{d} x^{\mu} \mathrm{d} x^{\nu}
  & = & a^2  (- 2 \delta_{ij} \Psi^{(2)} + 2 \partial_i \partial_j
  E^{(2)} \nonumber\\
  & & + \partial_i C_j^{(2)} + \partial_j
  C_i^{(2)} + H_{ij}^{(2)}) \mathrm{d} x^i \mathrm{d} x^j, 
\end{eqnarray}
The explicit expressions of the gauge invariant second order variables,
$E^{(2)}$, $\Psi^{(2)}$, $C_i^{(2)}$ and $H_{ij}^{(2)}$ are
presented in Eqs.~(\ref{84c})--(\ref{90}) with the expression of
$\mathcal{X}_{\mu \nu}^S$ in Appendix \ref{D}. Based on the temporal components of the
second order Einstein field equations and Eqs.~(\ref{91-0}),
(\ref{91}),(\ref{101})--(\ref{144c}), the gauge invariant second order 
 density perturbation is obtained to be
\end{multicols}
\hspace{0mm}\vspace{5mm}\ruleup
\begin{eqnarray}
  \rho^{(\rm{GI}, 2)} & = & - \frac{1}{3 (1 + w) \kappa a^2 \mathcal{H}^2}
  (8 \partial_i \partial_0 \Psi^{(1)} \partial^i \partial_0 \Psi^{(1)}
   \nonumber\\
  &  & + 6 (1 + w) \mathcal{H}^3 (3 \partial_0 \Psi^{(2)} - \Delta \partial_0
  E^{(2)} - 4 \partial_0 \Psi^{(1)} \Delta E^{(1)} + 4
  \Psi^{(1)} (3 \partial_0 \Psi^{(1)} - \Delta \partial_0 E^{(1)}) \nonumber\\
  &  & +
  4 \partial_i \partial_j \partial_0 E^{(1)} \partial^i \partial^j
  E^{(1)}) - 3 (1 + w) \mathcal{H}^2 (6 (\partial_0 \Psi^{(1)})^2 - 4 \partial_0
  \Psi^{(1)} \Delta \partial_0 E^{(1)} \nonumber\\
  &  & + 16 \Psi^{(1)} \Delta
  \Psi^{(1)} + 2 \Delta \Psi^{(2)} + 6 \partial_i \Psi^{(1)} \partial^i
  \Psi^{(1)} - 4 \partial_i \Delta E^{(1)} \partial^i \Psi^{(1)}
   \nonumber\\
  &  & + \Delta \partial_0 E^{(1)} \Delta \partial_0
  E^{(1)} - 4 \Delta E^{(1)} \Delta \Psi^{(1)} - \partial_i
  \Delta E^{(1)} \partial^i \Delta E^{(1)} - 4 \partial_i
  \partial_j \Psi^{(1)} \partial^i \partial^j E^{(1)} \nonumber\\
  &  &   - \partial_i \partial_j \partial_0
  E^{(1)} \partial^i \partial^j \partial_0 E^{(1)} +
  \partial_i \partial_j \partial_k E^{(1)} \partial^i \partial^j
  \partial^k E^{(1)})) \label{158c}.
\end{eqnarray}
\vspace{5mm}\ruledown
\begin{multicols}{2}\hspace{-6.5mm}
Substituting Eqs.~(\ref{91-0}), (\ref{91}),(\ref{101})--(\ref{144c}) and (\ref{158c}) into the spatial parts of the gauge invariant second order Einstein field equations, we also obtain
\begin{equation}
  \mathcal{G}_{i  j} +\mathcal{S}_{i  j} = 0,
\end{equation}
where $\mathcal{G}_{i  j}$ is shown in Eq.~(\ref{120c})
and $\mathcal{S}_{i  j}$ in Eq.~(\ref{149c}). Therefore, we
derive the equations of motion of the gauge invariant  second order cosmological
perturbations as
\end{multicols}
\hspace{0mm}\vspace{5mm}\ruleup
\begin{subequations}
\begin{eqnarray}
  \partial_0^2 H_{i j}^{(2)} + 2\mathcal{H} \partial_0 H_{i j}^{(2)} - \Delta H_{i j}^{(2)} & = & - 4
  \Lambda^{kl}_{ij} \mathcal{S}_{kl}, \label{eq:sai158}\\
  (\partial_0 + 2\mathcal{H}) \partial_0 C_i^{(2)} & = & - 4
  \Delta^{- 1} \mathcal{T}^k_i \partial^l \mathcal{S}_{k  l}, \label{eq:sai159}\\
  (\partial_0 + 2\mathcal{H}) \partial_0 E^{(2)} + \Psi^{(2)} & = & -
  2 \Delta^{- 1} \big( \partial^i \Delta^{- 1} \partial^j - \frac{1}{2}
  \mathcal{T}^{i  j} \big) \mathcal{S}_{i  j}, \\
  \partial_0^2 \Psi^{(2)} + (2 + 3 (c_s^{(2)})^2) \mathcal{H} \partial_0
  \Psi^{(2)} - \frac{1}{2} (1 + 2 (c_s^{(2)})^2) \Delta \Psi^{(2)} &  & 
  \nonumber\\
  - \Delta \big( \frac{1}{2} (\partial_0 + 2\mathcal{H}) \partial_0
  E^{(2)} + (c_s^{(2)})^2 \mathcal{H} \partial_0 E^{(2)}
  \big) & = & - \frac{1}{2} \mathcal{T}^{i  j} \mathcal{S}_{i
   j} . 
\end{eqnarray}\end{subequations}
\vspace{5mm}\ruledown
\begin{multicols}{2}\hspace{-6.5mm}
The right hand sides of the above equations are as same as that in Eqs.~(\ref{eq:sai151})--(\ref{eq:sai154}), since the source term $\mathcal{S}_{i  j}$ is uniquely determined by the gauge invariant first order synchronous variables. 
In particular, the equation of motion of the tensor perturbation $H_{\mu \nu}^{(\rm{GI},2)}$ in Eq.~(\ref{eq:sai158}) is as same as that in Eq.~(\ref{eq:sai147}). 
For the vector perturbation, the evolution of $\partial_0 C_i^{(2)}$ in Eq.~(\ref{eq:sai159}) is shown to be
as same as that of $V_i^{(2)}$ in Eq.~(\ref{eq:sai148}).

\section{Conclusions and discussions}\label{VI}

In this paper, we investigated the gauge invariance of the cosmological perturbations up to second order by following the Lie derivative method. 
We showed that there are infinite families of gauge invariant variables for the cosmological perturbations up to second order.
For different families, we found their conversion formulae which belong to a linear space spanned by a finite number of bases that were also shown to be gauge invariant. 
In particular, we have focused on the Newtonian, synchronous, and hybrid variables, respectively. We presented the explicit conversions between these different families of gauge-invariant variables.
In contrast to the first order, the second order gravitational waves were shown to be mixed with the first order cosmological perturbations. 
Therefore, the gauge invariance is important in studies of the second order gravitational waves.

We derived the equations of motion of the gauge invariant second order cosmological perturbations, which are sourced from the gauge invariant first order scalar perturbations. It was found that the choices of gauge-invariant variables at different orders are independent,
%On the one hand, the gauge invariant second order variables can take the same form as the gauge invariant first order variables. 
%On the other hand, it is not forbidden that they take different forms, 
e.g., Newtonian at first order while synchronous at second order, and vice versa. 
In this work, we have studied both of the above four typical cases. 
To obtain the gauge invariant equations of motion, we decomposed the gauge invariant perturbed Einstein field equations into the scalar, vector and tensor components. 

In principle, one could generalize the above method to explore the higher order cosmological perturbations. 
A formal derivation of the gauge invariant higher order cosmological perturbations has been shown previously {\cite{Nakamura:2014kza}}. 
Following the same approach in this work, it is straightforward to obtain the explicit expressions for the gauge invariant higher order scalar, vector and tensor perturbations, as well as the equations of motion of them. %We leave such a study for future works, since the derivations are much more complicated than here. 

This study is based on the method developed by \cite{Bruni:1996im,Nakamura:2006rk}. We further focus on conversion between different gauge-invariant variables. Since there is no uniqueness in the framework, it might show that one can not determine which gauge-invariant tensor perturbations $H^{(2)}_{i j}$ correspond to the energy density spectrum of gravitational waves. Perhaps, physical insights or arguments should be explored in future studies.

\acknowledgments
We acknowledge Prof. Rong-Gen Cai, Prof. Qing-Guo Huang, Prof. Xin Li, Prof. Tao Liu, Prof. Shi Pi, Prof. Jiang-Hao Yu, Prof. Xin Zhang, Mr. Zu-Cheng Chen, Mr. Chen Yuan, Mr. Xukun Zhang and Mr. Jing-Zhi Zhou for helpful discussions. 
This work is supported by the National Natural Science Foundation of China upon Grant No. 12075249, No. 11675182 and No. 11690022, and by a grant upon Grant No. Y954040101 from the Institute of High Energy Physics, Chinese Academy of Sciences. We acknowledge the \texttt{xPand} package \cite{Pitrou:2013hga}.

\appendix

\section{Expansion of a generic tensor with Lie derivative}\label{A}

For a generic tensor $\mathcal{Q}$, its Lie derivative along a vector $\zeta^{\mu}$ is defined as
\begin{equation}
  \mathcal{L}_{\zeta} \mathcal{Q} \equiv \lim_{\epsilon \rightarrow 0} 
  \frac{\varphi^{\ast}_{\epsilon} \mathcal{Q} (x) -\mathcal{Q} (x)}{\epsilon},
  \label{107}
\end{equation}
where $\varphi^{\ast}_{\epsilon} \mathcal{Q}$ is an one-parameter coordinate
transformation that acts on the tensor $\mathcal{Q}$. In the regime of
$\epsilon \rightarrow 0$, it turns to be the infinitesimal
transformation in the form of Eq.~(\ref{1}), where $\xi^{(1) \mu} \propto
\epsilon \zeta^{\mu}$. For a scalar function, contravariant vector and covariant vector, the
expressions of $\varphi_{\epsilon}^{\ast}$ are given by
\begin{subequations}
\begin{eqnarray}
  \varphi^{\ast}_{\epsilon} f (x) & = & f (\tilde{x}),  \label{108}\\
  \varphi^{\ast}_{\epsilon} w_{\mu} (x) & = & \frac{\partial
  \tilde{x}^{\nu}}{\partial x^{\mu}} w_{\nu} (\tilde{x}), \\
  \varphi_{\epsilon, \ast} A^{\mu} (x) & = & \frac{\partial x^{\mu}}{\partial
  \tilde{x}^{\nu}} A^{\nu} (\tilde{x}) .  \label{110}
\end{eqnarray}\end{subequations}
Based on the first order expansions of Eqs.~(\ref{108})--(\ref{110}), we obtain 
\begin{subequations}\begin{eqnarray}
  \mathcal{L}_{\zeta} f & = & \zeta^{\nu} \partial_{\nu} f, \\
  \mathcal{L}_{\zeta} w_{\mu} & = & \zeta^{\nu} \partial_{\nu} w_{\mu} +
  w_{\nu} \partial_{\mu} \zeta^{\mu}, \\
  \mathcal{L}_{\zeta} A^{\mu} & = & \zeta^{\nu} \partial_{\nu} A^{\mu} -
  A^{\nu} \partial_{\nu} \zeta^{\mu} .  \label{113}
\end{eqnarray}\end{subequations}
%The derivations can be found in standard textbook. 
The Lie derivative in Eq.~(\ref{113}) is also symbolized as $\mathcal{L}_X A^{\mu} \equiv [X,
A]^{\mu}$, where $[,]$ is Lie bracket. For a tensor $S_{\mu \nu}$, its Lie
derivative is given by
\begin{equation}
  \mathcal{L}_{\zeta} S_{\mu \nu} = \zeta^{\lambda} \partial_{\lambda} S_{\mu
  \nu} + S_{\lambda \nu} \partial_{\mu} \zeta^{\lambda} + S_{\mu \lambda}
  \partial_{\nu} \zeta^{\lambda} .
\end{equation}
Higher order expansions of any tensor upon the infinitesimal transformation have been constructed in terms of Lie derivatives {\cite{Matarrese:1997ay,Sonego:1997np,Nakamura:2006rk,Malik:2008im,Bruni:1996im}}. 
In the following, we briefly review the formulae up to second order.

For the scalar function $f (x)$ upon the infinitesimal transformation, up to second order, we expand it in terms of $\xi^{(1)}$ and $\xi^{(2)}$, namely, 
\begin{eqnarray}
  f (x) & \rightarrow & f (\tilde{x}) \nonumber\\
  & \approx & f \big( x + \xi^{(1)} + \frac{1}{2} \xi^{(2)} \big)
  \nonumber\\
  & = & f + \xi^{(1) \nu} \partial_{\nu} f + \frac{1}{2} \xi^{(2) \nu}
  \partial_{\nu} f + \frac{1}{2} \xi^{(1) \sigma} \xi^{(1) \rho}
  \partial_{\sigma} \partial_{\rho} f \nonumber\\
  & = & f + \xi^{(1) \nu} \partial_{\nu} f + \frac{1}{2}  (\xi^{(2) \nu}
  \partial_{\nu} f + \xi^{(1) \sigma} \xi^{(1) \rho} \partial_{\sigma}
  \partial_{\rho} f \nonumber\\
  & &  + \xi^{(1) \sigma} \partial_{\sigma} \xi^{(1) \rho}
  \partial_{\rho} f - \xi^{(1) \sigma} \partial_{\sigma} \xi^{(1) \rho}
  \partial_{\rho} f) \nonumber\\
  & = & f +\mathcal{L}_{\xi^{(1)}} f + \frac{1}{2}  (\mathcal{L}_{\xi^{(2)} -
  \xi^{(1) \nu} \partial_{\nu} \xi^{(1)}} +\mathcal{L}^2_{\xi^{(1)}}) f.   \label{A9}
\end{eqnarray}
For simplicity, we denote
\begin{subequations}
\begin{eqnarray}
  \xi_1^{\mu} & \equiv & \xi^{(1) \mu}, \\
  \xi_2^{\mu} & \equiv & \xi^{(2) \mu} - \xi^{(1) \nu} \partial_{\nu} \xi^{(1)
  \mu} . 
\end{eqnarray}
\end{subequations}
It leads to
\begin{equation}
  f (\tilde{x}) = f +\mathcal{L}_{\xi_1} f + \frac{1}{2}  (\mathcal{L}_{\xi_2}
  +\mathcal{L}^2_{\xi_1}) f +\mathcal{O} (\xi^{(3)}) .
\end{equation}
For the contravariant vector $w_{\mu}$, up to second order, we expand it in terms of $\xi^{(1)}$ and $\xi^{(2)}$, namely, 
\begin{eqnarray}
  w_{\mu} (x) & \rightarrow & \frac{\partial \tilde{x}^{\nu}}{\partial
  x^{\mu}} w_{\nu} (\tilde{x}) \nonumber\\
  & \approx & \big( \delta^{\lambda}_{\mu} + \partial_{\mu} \xi^{(1)
  \lambda} + \frac{1}{2} \partial_{\mu} \xi^{(2) \lambda} \big)  \Big(
  w_{\lambda} (x) \nonumber\\
  & &  + \big( \xi^{(1) \nu} + \frac{1}{2} \xi^{(2) \nu} \big)
  \partial_{\nu} w_{\lambda}  \nonumber\\
  &  &  + \big( \xi^{(1) \sigma} + \frac{1}{2} \xi^{(2) \sigma}
  \big)  \big( \xi^{(1) \rho} + \frac{1}{2} \xi^{(1) \rho} \big)
  \partial_{\sigma} \partial_{\rho} w_{\lambda} \Big) \nonumber\\
  & = & w_{\mu} (x) + \xi^{(1) \nu} \partial_{\nu} w_{\mu} + w_{\lambda}
  \partial_{\mu} \xi^{(1) \lambda} +
  \frac{1}{2} w_{\lambda} \partial_{\mu} \xi^{(2) \lambda} \nonumber\\
  & &  + \partial_{\mu} \xi^{(1) \lambda} \xi^{(1)
  \nu} \partial_{\nu} w_{\lambda} \nonumber\\
  &  & + \frac{1}{2} ((\xi^{(2) \nu} \partial_{\nu} w_{\mu} + \xi^{(1)
  \sigma} \xi^{(1) \rho} \partial_{\sigma} \partial_{\rho} w_{\mu}) \nonumber\\
  & = & w_{\mu} +\mathcal{L}_{\xi^{(1)}} w_{\mu} + \frac{1}{2}
  \mathcal{L}_{\xi^{(2)}} w_{\mu} \nonumber\\
  & & + \frac{1}{2} (\mathcal{L}_{\xi^{(1)}}^2
  w_{\mu} - \xi^{(1) \sigma} \partial_{\sigma} \xi^{(1) \rho} \partial_{\rho}
  w_{\mu} \nonumber\\
  & &  - w_{\rho} \partial_{\mu} ((\xi^{(1) \sigma} \partial_{\sigma}
  \xi^{(1) \rho})) \nonumber\\
  & = & w_{\mu} +\mathcal{L}_{\xi^{(1)}} w_{\mu} + \frac{1}{2} 
  (\mathcal{L}_{\xi^{(2)} - \xi^{(1) \sigma} \partial_{\sigma} \xi^{(1)}}
  +\mathcal{L}_{\xi^{(1)}}^2) w_{\mu} \nonumber\\
  & = & w_{\mu} +\mathcal{L}_{\xi_1} w_{\mu} + \frac{1}{2} 
  (\mathcal{L}_{\xi_2} +\mathcal{L}_{\xi_1}^2) w_{\mu} .  \label{A13}
\end{eqnarray}
Similarly, for the covariant vector $A^{\mu}$, up to second order, we expand it in terms of $\xi^{(1)}$ and $\xi^{(2)}$, namely, 
\begin{eqnarray}
  A^{\mu} (x) & \rightarrow & \frac{\partial x^{\mu}}{\partial
  \tilde{x}^{\nu}} A^{\nu} (\tilde{x}) \nonumber\\
  & \approx &  \left( \frac{\partial \tilde{x}^{\mu}}{\partial x^{\nu}}
  \right)^{- 1}  A^{\nu}  \big( x + \xi^{(1)} + \frac{1}{2} \xi^{(2)}
  \big) \nonumber\\
  & \approx & A^{\mu} + \xi^{(1) \sigma} \partial_{\sigma} A^{\mu} - A^{\nu}
  \partial_{\nu} \xi^{(1) \mu} - \partial_{\nu} \xi^{(1) \mu} \xi^{(1) \sigma}
  \partial_{\sigma} A^{\nu} \nonumber\\
  & & - \frac{1}{2}  (\partial_{\nu} \xi^{(2) \mu} - 2
  \partial_{\lambda} \xi^{(1) \mu} \partial_{\nu} \xi^{(1) \lambda}) A^{\nu}
  \nonumber\\
  &  & + \frac{1}{2} ((\xi^{(2) \sigma} \partial_{\sigma} A^{\mu} + \xi^{(1)
  \sigma} \xi^{(1) \rho} \partial_{\sigma} \partial_{\rho} A^{\mu})
  \nonumber\\
  & = & A^{\mu} +\mathcal{L}_{\xi^{(1)}} A^{\mu} + \frac{1}{2} 
  (\mathcal{L}_{\xi^{(2)} - \xi^{(1) \nu} \partial_{\nu} \xi^{(1)})} A^{\mu}
  +\mathcal{L}^2_{\xi^{(1)}} A^{\mu})) \nonumber\\
  & = & A^{\mu} +\mathcal{L}_{\xi_1} A^{\mu} + \frac{1}{2} 
  (\mathcal{L}_{\xi_2} A^{\mu} +\mathcal{L}^2_{\xi_1} A^{\mu}), \label{A14}
\end{eqnarray}
where the inverse Jacobi matrix ${\partial x^{\mu}}/{\partial
\tilde{x}^{\nu}}$ is given by
\begin{equation}
  \left( \frac{\partial \tilde{x}^{\mu}}{\partial x^{\nu}} \right)^{- 1} =
  \delta^{\mu}_{\nu} - \partial_{\nu} \xi^{(1) \mu} - \frac{1}{2} 
  (\partial_{\nu} \xi^{(2) \mu} - 2 \partial_{\lambda} \xi^{(1) \mu}
  \partial_{\nu} \xi^{(1) \lambda}) +\mathcal{O} (\xi^{(3)}) .
\end{equation}
It is derived from $\left( \frac{\partial \tilde{x}^{\lambda}}{\partial
x^{\nu}} \right)^{- 1} \frac{\partial \tilde{x}^{\mu}}{\partial x^{\lambda}} =
\frac{\partial x^{\lambda}}{\partial \tilde{x}^{\nu}}  \frac{\partial
\tilde{x}^{\mu}}{\partial x^{\lambda}} = \delta^{\mu}_{\nu}$. Finally, we
consider the tensor $S_{\mu \nu}$ which is expanded as 
\end{multicols}
\hspace{0mm}\vspace{5mm}\ruleup
\begin{eqnarray}
  S_{\sigma \rho} (x) & \rightarrow & \frac{\partial \tilde{x}^{\mu}}{\partial
  x^{\sigma}}  \frac{\partial \tilde{x}^{\nu}}{\partial x^{\rho}} S_{\mu \nu}
  (\tilde{x}) \nonumber\\
  & \approx & S_{\mu \nu}  \big( x + \xi^{(1)} + \frac{1}{2} \xi^{(2)}
  \big) \partial_{\sigma}  \big( x^{\mu} + \xi^{(1) \mu} + \frac{1}{2}
  \xi^{(2) \mu} \big) \partial_{\rho}  \big( x^{\nu} + \xi^{(1) \nu} +
  \frac{1}{2} \xi^{(2) \nu} \big) \nonumber\\
  & \approx & S_{\sigma \rho} + \xi^{(1) \lambda} \partial_{\lambda}
  S_{\sigma \rho} + S_{\sigma \nu} \partial_{\rho} \xi^{(1) \nu} + S_{\mu
  \rho} \partial_{\sigma} \xi^{(1) \mu} + \frac{1}{2} ((\xi^{(2) \lambda}
  \partial_{\lambda} S_{\sigma \rho} + S_{\mu \rho} \partial_{\sigma} \xi^{(2)
  \mu} + S_{\sigma \nu} \partial_{\rho} \xi^{(2) \nu} \nonumber\\
  &  & + \xi^{(1) \lambda} \xi^{(1) \kappa} \partial_{\lambda}
  \partial_{\kappa} S_{\sigma \rho} + 2 S_{\mu \nu} \partial_{\sigma} \xi^{(1)
  \mu} \partial_{\rho} \xi^{(1) \nu} + 2 \xi^{(1) \lambda} \partial_{\lambda}
  S_{\sigma \nu} \partial_{\rho} \xi^{(1) \nu} + 2 \xi^{(1) \lambda}
  \partial_{\lambda} S_{\mu \rho} \partial_{\sigma} \xi^{(1) \mu}) \nonumber\\
  & = & S_{\sigma \rho} +\mathcal{L}_{\xi^{(1)}} S_{\sigma \rho} +
  \frac{1}{2}  (\mathcal{L}_{\xi^{(2)} - \xi^{(1) \nu} \partial_{\nu}
  \xi^{(1)}} +\mathcal{L}_{\xi^{(1)}}^2) S_{\sigma \rho} \nonumber\\
  & = & S_{\sigma \rho} +\mathcal{L}_{\xi_1} S_{\sigma \rho} + \frac{1}{2} 
  (\mathcal{L}_{\xi_2} +\mathcal{L}_{\xi_1}^2) S_{\sigma \rho} . \label{A16}
\end{eqnarray}\vspace{5mm}\ruledown
\begin{multicols}{2}\hspace{-6.5mm}
The above formulae can be straightforwardly generalized to higher order expansions. 
%In the same way, it might not be difficult to extend the derivations for
%higher order expansions of the Lie derivative.

In these derivations, we use the Eq.~(\ref{1}) for expanding the $\tilde{x}$. It can be obtained via
\begin{equation}
	\tilde{x}^\mu \equiv \varphi_\epsilon x^\mu = x^\mu + \frac{{\rm d}\varphi_\epsilon x^\mu}{{\rm d} \epsilon}\Big|_{\epsilon=0}\epsilon + \frac{1}{2}\frac{{\rm d}^2\varphi_\epsilon x^\mu}{{\rm d} \epsilon^2}\Big|_{\epsilon=0}\epsilon^2 + \mathcal{O}(\epsilon^3)~, \label{A106}
\end{equation}
where the explicit expressions of $\xi^{(1)}$ and $\xi^{(2)}$ are defined as
\begin{eqnarray}
	\xi^{(1)}\equiv\frac{{\rm d}\varphi_\epsilon x^\mu}{{\rm d} \epsilon}\Big|_{\epsilon=0}\epsilon~, \\
	\xi^{(2)}\equiv \frac{{\rm d}^2\varphi_\epsilon x^\mu}{{\rm d} \epsilon^2}\Big|_{\epsilon=0}\epsilon^2~.
\end{eqnarray}
Therefore, the order of the $\xi^{(n)}$ is determined by power of $\epsilon$ in the transformation (Eq.~(\ref{A106})), and has no relevance with metric perturbations or matter perturbations in principle.

\section{Gauge transformations in the language of Lie derivative}\label{A-1}

For an arbitrary tensor $\mathcal{Q} (x^{\mu})^{i_1 i_2 \ldots}_{j_1 j_2
\ldots}$ upon the infinitesimal transformation, it can be
expanded as \cite{Bruni:1996im}
\end{multicols}
\hspace{0mm}\vspace{5mm}\ruleup
\begin{eqnarray}
  \mathcal{Q} (x)^{i_1 i_2 \ldots}_{j_1 j_2 \ldots} & \rightarrow & \big(
    \frac{\partial \tilde{x}^{l_1}}{\partial
  {x}^{j_1}}  \frac{\partial \tilde{x}^{l_2}}{\partial {x}^{j_2}} \frac{\partial {x}^{i_1}}{\partial \tilde{x}^{k_1}}  \frac{\partial
  {x}^{i_2}}{\partial \tilde{x}^{k_2}} \ldots
  \big) \mathcal{Q} (\tilde{x})^{k_1 k_2 \ldots}_{l_1 l_2 \ldots}
  \nonumber\\
  & \approx & \big( \delta^{l_1}_{j_1} + \partial_{j_1} \xi^{(1) l_1} +
  \frac{1}{2} \partial_{j_1} \xi^{(2) l_1} \big) \big(\ldots\big) \Big(\mathcal{Q}
  (x)^{k_1 k_2 \ldots}_{l_1 l_2 \ldots} + \xi^{(1) \nu} \partial_{\nu}
  \mathcal{Q} (x)^{k_1 k_2 \ldots}_{l_1 l_2 \ldots} \nonumber\\
  &  &  + \frac{1}{2} \big((\xi^{(1) \sigma} \xi^{(2) \rho}
  \partial_{\sigma} \partial_{\rho} \mathcal{Q}(x)^{k_1 k_2 \ldots}_{l_1 l_2
  \ldots} + \xi^{(2) \nu} \partial_{\nu} \mathcal{Q}(x)^{k_1 k_2 \ldots}_{l_1
  l_2 \ldots}) \big) \Big)  \nonumber\\
  & \approx & \mathcal{Q} (x^{\mu})^{i_1 i_2 \ldots}_{j_1 j_2 \ldots}
  +\mathcal{L}_{\xi^{(1)}} \mathcal{Q} (x^{\mu})^{i_1 i_2 \ldots}_{j_1 j_2
  \ldots} + \frac{1}{2}  (\mathcal{L}_{\xi^{(2)} - \xi^{(1) \nu}
  \partial_{\nu} \xi^{(1)}} +\mathcal{L}_{\xi^{(1)}}^2) \mathcal{Q}
  (x^{\mu})^{i_1 i_2 \ldots}_{j_1 j_2 \ldots}  \label{2}
\end{eqnarray}\vspace{5mm}\ruledown
\begin{multicols}{2}\hspace{-6.5mm}
For simplicity, the above equation can be rewritten without indices as follows
%we could set $\xi_1 \equiv \xi^{(1)}$, $\xi_2 \equiv \xi^{(2)}
%- \xi^{(1) \nu} \partial_{\nu} \xi^{(1)}$ and rewrite the tensor $\mathcal{Q}
%(x^{\mu})^{i_1 i_2 \ldots}_{j_1 j_2 \ldots}$ without indices,
\begin{eqnarray}
  \mathcal{Q} \rightarrow \tilde{\mathcal{Q}}  = 
  \mathcal{Q}+\mathcal{L}_{\xi_1} \mathcal{Q}+ \frac{1}{2} 
  (\mathcal{L}_{\xi_2} +\mathcal{L}_{\xi_1}^2) \mathcal{Q}+\mathcal{O}
  (\xi^{(3)})  \label{3}
\end{eqnarray}
In Appendix \ref{A}, we have presented four examples of Eq.~(\ref{3}), namely, Eqs.~(\ref{A9}), (\ref{A13}), (\ref{A14}) and (\ref{A16}). 
%Formal proof of Eq.~(\ref{3})was given by Bruni {\cite{bruni_perturbations_1997}} through introducing the onception of knight diffeomorphisms.

In the following, we consider the infinitesimal transformation of perturbations up to second order. The $n$-th order perturbations $\mathcal{Q}^{(n)}$ of the tensor $\mathcal{Q}$ can be expanded as follows
\begin{eqnarray}
  \mathcal{Q}  =  \mathcal{Q}^{(0)} +\mathcal{Q}^{(1)} + \frac{1}{2}
  \mathcal{Q}^{(2)} +\mathcal{O} (\mathcal{Q}^{(3)}) .  \label{4}
\end{eqnarray}
In order to study the gauge transformation of $\mathcal{Q}^{(n)}$, one can choose a type of infinitesimal transformation where the order of $\xi^{(1)}$ is the same order of $\mathcal{Q}^{(1)}$. 
Combining Eqs. (\ref{3}) with (\ref{4}), we obtain
\begin{eqnarray}
  \tilde{\mathcal{Q}} &=&\mathcal{Q}^{(0)} + (\mathcal{Q}^{(1)}
  +\mathcal{L}_{\xi_1} \mathcal{Q}^{(0)}) \nonumber\\
  & & + \frac{1}{2}  (\mathcal{Q}^{(2)} +
  2\mathcal{L}_{\xi_1} \mathcal{Q}^{(1)} + (\mathcal{L}_{\xi_2}
  +\mathcal{L}_{\xi_1}^2)\mathcal{Q}^{(0)}) \nonumber\\
  & & +\mathcal{O} (\xi^{(3)}) \label{5}
  .
\end{eqnarray}
On the other side, the tensor $\tilde{\mathcal{Q}}$ can also be
expanded in the form of Eq. (\ref{4}), namely,
\begin{equation}
  \tilde{\mathcal{Q}} = \tilde{\mathcal{Q}}^{(0)} + \tilde{\mathcal{Q}}^{(1)}
  + \frac{1}{2}  \tilde{\mathcal{Q}}^{(2)} +\mathcal{O} (\tilde{\mathcal{Q}}^{(3)} ) .
  \label{5-1}
\end{equation}
Therefore, based on Eqs.~(\ref{5}) and (\ref{5-1}), we conclude the infinitesimal
transformations of the zeroth, first and second order perturbations in Eqs.~(\ref{A1}).

\section{Brief derivation of Eqs. (\ref{13}) and
(\ref{14})}\label{B}

For the first order metric perturbations $g^{(1)}_{\mu \nu}$, it could be
split into a gauge invariant part $g^{(\mathrm{\rm{GI}, 1)}}_{\mu \nu}$
and a gauge variant counter term $\mathbb{C}_{\mu \nu}^{(1)}$, i.e.,
\begin{eqnarray}
  g_{\mu \nu}^{(1)} & = : & g_{\mu \nu}^{(\mathrm{\rm{GI}, 1)}} + \mathbb{C}_{\mu
  \nu}^{(1)} .  \label{123}
\end{eqnarray}
Based on the gauge 
transformation of $g_{\mu \nu}^{(1)}$ in Eq. (\ref{9}), we could rewrite
$g^{(\mathrm{\rm{GI}, 1)}}_{\mu \nu}$ to be
\begin{eqnarray}
  g^{(\mathrm{\rm{GI}, 1)}}_{\mu \nu} & = & \tilde{g}_{\mu
  \nu}^{(\mathrm{\rm{GI}, 1)}} \nonumber\\
  & = & \tilde{g}_{\mu \nu}^{(1)} - \tilde{\mathbb{C}}_{\mu \nu}^{(1)} \nonumber\\
  & = & g_{\mu \nu}^{(1)} +\mathcal{L}_{\xi_1} g_{\mu \nu}^{(0)} -
  \tilde{\mathbb{C}}_{\mu \nu}^{(1)} .  \label{124}
\end{eqnarray}
Based on Eqs. (\ref{123}) and (\ref{124}), we obtain
\begin{equation}
  \tilde{\mathbb{C}}^{(1)}_{\mu \nu} = \mathbb{C}^{(1)}_{\mu \nu} +\mathcal{L}_{\xi_1} g_{\mu
  \nu}^{(0)} . \label{125}
\end{equation}
If we let $\mathbb{C}^{(1)}_{\mu \nu} \equiv \mathcal{L}_X g_{\mu \nu}^{(0)}$, Eq.
(\ref{125}) is reduced to $\mathcal{L}_{\tilde{X}} g_{\mu \nu}^{(0)}
=\mathcal{L}_{X + \xi_1} g_{\mu \nu}^{(0)}$. 
We can obtain
\begin{equation}
  \tilde{X}^{\mu} - X^{\mu} = \xi_1^{\mu} .
\end{equation}
Here, $X^\mu$ is rewritten in the form independent on the background metric.
Therefore, we can rewrite Eq. (\ref{123}) in the form of
\begin{equation}
  g_{\mu \nu}^{(1)} = g_{\mu \nu}^{(\mathrm{\rm{GI}, 1)}} +\mathcal{L}_X
  g_{\mu \nu}^{(0)} .
\end{equation}
For the second order metric perturbations $g_{\mu \nu}^{(2)}$, we could also
split it as
\begin{equation}
  g_{\mu \nu}^{(2)} = : g_{\mu \nu}^{(\mathrm{\rm{GI}, 2)}} + \mathbb{C}_{\mu
  \nu}^{(2)}, \label{128}
\end{equation}
where $g_{\mu \nu}^{(\mathrm{\rm{GI}, 2)}}$ are the gauge invariant second order
metric perturbations. 
Based on Eqs. (\ref{10}) and (\ref{128}), we can also express $g_{\mu
\nu}^{(\mathrm{\rm{GI}, 2)}}$ as
\begin{eqnarray}
  g_{\mu \nu}^{(\mathrm{\rm{GI}, 2)}} & = & \tilde{g}_{\mu
  \nu}^{(\mathrm{\rm{GI},2)}} \nonumber\\
  & = & \tilde{g}_{\mu \nu}^{(2)} - \tilde{\mathbb{C}}_{\mu \nu}^{(2)} \nonumber\\
  & = & 2\mathcal{L}_{\xi_1} g_{\mu \nu}^{(1)} + (\mathcal{L}_{\xi_2}
  +\mathcal{L}_{\xi_1}^2) g_{\mu \nu}^{(0)} - \tilde{\mathbb{C}}_{\mu \nu}^{(2)} . 
  \label{129}
\end{eqnarray}
Based on Eqs. (\ref{128}) and (\ref{129}), we have
\begin{eqnarray}
  \tilde{\mathbb{C}}_{\mu \nu}^{(2)} - \mathbb{C}_{\mu \nu}^{(2)} & = & 2\mathcal{L}_{\xi_1}
  g_{\mu \nu}^{(1)} + (\mathcal{L}_{\xi_2} +\mathcal{L}_{\xi_1}^2) g_{\mu
  \nu}^{(0)} \nonumber\\
  & = & 2\mathcal{L}_{\tilde{X}}  (\tilde{g}_{\mu \nu}^{(1)}
  -\mathcal{L}_{\xi_1} g_{\mu \nu}^{(0)}) - 2\mathcal{L}_X g_{\mu \nu}^{(1)}
  +\mathcal{L}_{\xi_2} g_{\mu \nu}^{(0)} \nonumber\\
  & &+ (\mathcal{L}_{\tilde{X}}^2
   +\mathcal{L}_X^2 -\mathcal{L}_{\tilde{X}} \mathcal{L}_X -\mathcal{L}_X
  \mathcal{L}_{\tilde{X}}) g_{\mu \nu}^{(0)} \nonumber\\
  & = & (2\mathcal{L}_{\tilde{X}}  \tilde{g}_{\mu \nu}^{(1)} -
  \mathcal{L}^2_{\tilde{X}} g_{\mu \nu}^{(0)}) - (2\mathcal{L}_X g_{\mu \nu}^{(1)}
  \nonumber\\
  & & -\mathcal{L}_X^2 g_{\mu \nu}^{(0)}) +\mathcal{L}_{(\xi_2 + [\tilde{X}, X])}
  g_{\mu \nu}^{(0)} \nonumber\\
  & \equiv & (2\mathcal{L}_{\tilde{X}}  \tilde{g}_{\mu \nu}^{(1)} +
  (\mathcal{L}_{\tilde{Y}} - \mathcal{L}^2_{\tilde{X}}) g_{\mu \nu}^{(0)}) \nonumber\\
  & & -
  (2\mathcal{L}_X g_{\mu \nu}^{(1)} + (\mathcal{L}_Y -\mathcal{L}_X^2) g_{\mu
  \nu}^{(0)}), 
\end{eqnarray}
where we have introduced the infinitesimal vector $Y^{\mu}$ satisfying 
\begin{equation}
  \tilde{Y}^{\mu} - Y^{\mu} = \xi_2^{\mu} + [\xi_1, X]^{\mu} .
\end{equation}
Therefore, we can rewrite Eq. (\ref{128}) in the form of
\begin{equation}
  g_{\mu \nu}^{(2)} = g^{(\mathrm{\rm{GI}, 2)}}_{\mu \nu} + 2\mathcal{L}_X
  g^{(1)}_{\mu \nu} + (\mathcal{L}_Y -\mathcal{L}_X^2) g_{\mu \nu}^{(0)} .
\end{equation}

\section{Scalar-vector-tensor decomposition}\label{C}

By making use of the transverse operators, namely,
$\mathcal{T}^i_j \equiv \delta^i_j - \partial^i \Delta^{- 1} \partial_j$, we could decompose an arbitrary spatial vector $U_i$ to be 
\begin{eqnarray}
  U_i & = & U^{(T)}_i + \partial_i U^{(S)}, 
\end{eqnarray}
where the transverse part $U_i^{(T)}$ and the longitudinal part $U^{(S)} $ are defined as
\begin{subequations}
\begin{eqnarray}
  U_i^{(T)} & \equiv & \mathcal{T}^k_i U_k, \\
  U^{(S)} & \equiv & \Delta^{- 1} \partial^i U_i . 
\end{eqnarray}
\end{subequations}
and $\Delta^{- 1}$ is the inverse Laplacian operator.

For an arbitrary spatial tensor $S_{ij}$, we could decompose it as \cite{Weinberg:2008zzc,Maggiore:2018sht}
\begin{eqnarray}
  S_{ij} & = & S^{(H)}_{ij} + 2 \delta_{ij} S^{(\Psi)} + 2 \partial_i
  \partial_j S^{(E)} + \partial_j S_i^{(C)} + \partial_i S_j^{(C)}, 
\end{eqnarray}
where we can define the tensor part $S^{(H)}_{ij}$, 
the vector part $S_i^{(C)}$  and the scalar parts $S^{(\Psi)}$, $S^{(E)}$ as 
\begin{subequations}
\begin{eqnarray}
  S^{(H)}_{ij} & \equiv & \big( \mathcal{T}^k_i \mathcal{T}^l_j - \frac{1}{2}
  \mathcal{T}_{ij} \mathcal{T}^{kl} \big) S_{kl}, \\
    S_i^{(C)} & \equiv & \Delta^{- 1} \partial^l \mathcal{T}^k_i S_{lk} , \\
  S^{(\Psi)} & \equiv & \frac{1}{4} \mathcal{T}^{kl} S_{kl}, \\
  S^{(E)} & \equiv & \frac{1}{2} \Delta^{- 1}  \big( \partial^k \Delta^{- 1}
  \partial^l - \frac{1}{2} \mathcal{T}^{kl} \big) S_{lk}.
\end{eqnarray}
\end{subequations}
Here, ($\mathcal{T}^k_i \mathcal{T}^l_j - \mathcal{T}_{ij}
\mathcal{T}^{kl}/2$) denotes the transverse and traceless operator. 
Other kinds of decomposition can be found in Refs.~\cite{Nakamura:2011xy,york_covariant_1974}. %except that it was traceless. 
%An other kind of decomposition can be found in Ref.~.

%In this work, we have used this decomposition for infinitesimal
%transformations of metric perturbations [Eq.~(\ref{56}) and (\ref{57})] and second order Einstein field equations [Eqs.~(\ref{124d})--(\ref{128c})]. The scalar, vector and tensor parts of the
%metric perturbations will not get entangled with other parts of the metric
%perturbations upon the infinitesimal transformations.

\section{Explicit expressions of $\mathcal{X}_{\mu \nu}$}\label{D}

For a given $X^{\mu}$, we express $\mathcal{X}_{\mu \nu}$ in terms
of the first order metric perturbations by following Eq.~(\ref{67c}).
The results are given by
\end{multicols}
\hspace{0mm}\vspace{5mm}\ruleup
\iffalse
\begin{eqnarray}
  \mathcal{X}_{00} & = & -2 \big( \frac{2 \dot{a}}{a} X^0 + X^{\mu}
  \partial_{\mu} + 2 \partial_0 X^0 \big)  \big( 2 \phi^{(1)} - (
  \partial_0 + \frac{\dot{a}}{a} ) X^0 \big) \nonumber\\
  &  &+ 2 \partial_0 X^i  (2
  \partial_i b^{(1)} + 2 \nu_i^{(1)} - \delta_{ji} \partial_0 X^j + \partial_i
  X^0), \\
  \mathcal{X}_{0 i} & = &  \big( \delta^k_j  ( \frac{2 \dot{a}}{a} X^0
  + X^{\mu} \partial_{\mu} + \partial_0 X^0 ) + \partial_j X^k \big) 
  (2 \partial_k b^{(1)} + 2 \nu_k^{(1)} - \delta_{lk} \partial_0 X^l +
  \partial_k X^0) \nonumber\\
  &  & + \partial_0 X^k  \big( - 4 \psi^{(1)} \delta_{jk} + 4 \partial_j
  \partial_k {e}^{(1)} + 2 \partial_j c_k^{(1)} + 2 \partial_k c_j^{(1)} + 2 h_{j k}^{(1)}  -
  (\delta_{jl} \partial_k + \delta_{kl} \partial_j) X^l - \frac{2 \dot{a}}{a}
  \delta_{kj} X^0 \big) \nonumber\\
  &  & - 2 \partial_j X^0  \big( 2 \phi^{(1)} - ( \partial_0 +
  \frac{\dot{a}}{a} ) X^0 \big), \\
  \mathcal{X}_{kl} & = &  \big( \delta^s_k \delta^t_l  ( \frac{2
  \dot{a}}{a} X^0 + X^{\mu} \partial_{\mu} ) + \delta^s_k \partial_l X^t
  + \delta^t_l \partial_k X^s \big) \nonumber\\
  &  &  \big( - 4 \psi^{(1)} \delta_{st} + 4 \partial_s \partial_t
  {e}^{(1)} + 2 \partial_t c_s^{(1)} + 2 \partial_s c_t^{(1)} + 2 h_{st}^{(1)} -
  (\delta_{sn} \partial_t + \delta_{tn} \partial_s) X^n - \frac{2 \dot{a}}{a}
  \delta_{st} X^0 \big) \nonumber\\
  &  & + (\delta^s_k \partial_l X^0 + \delta^s_l \partial_k X^0)  (2
  \partial_s b^{(1)} + 2 \nu_s^{(1)} - \delta_{ts} \partial_0 X^t + \partial_s
  X^0) . 
\end{eqnarray}
  \fi
\begin{eqnarray}
	\mathcal{X}_{\mu \nu} & = & \frac{4}{a^2}\mathcal{H}X^0 {g}^{(1)}_{\mu \nu} + 2
	X^{\sigma} \partial_{\sigma} \left(\frac{1}{a^2}{g}_{\mu \nu}^{(1)}\right) + \frac{2}{a^2} {g}_{\sigma
		\nu}^{(1)} \partial_{\mu} X^{\sigma} + \frac{2}{a^2} {g}_{\mu \sigma}^{(1)}
	\partial_{\nu} X^{\sigma}\nonumber\\
	&  & - \eta_{\mu \nu} ((4\mathcal{H}^2 + 2
	\dot{\mathcal{H}}) (X^0)^2 + 2\mathcal{H}X^{\mu} \partial_{\mu} X^0) - 2 \eta_{\sigma \rho}
	\partial_{\mu} X^{\sigma} \partial_{\nu} X^{\rho} \nonumber\\
	&  &-
	\eta_{\mu \sigma} (X^{\rho} \partial_{\rho} \partial_{\nu} X^{\sigma} +
	\partial_{\rho} X^{\sigma} \partial_{\nu} X^{\rho} + 4\mathcal{H}X^0
	\partial_{\nu} X^{\sigma}) \nonumber\\
	&  &- \eta_{\sigma \nu} (X^{\rho} \partial_{\rho}
	\partial_{\mu} X^{\sigma} + \partial_{\rho} X^{\sigma} \partial_{\mu}
	X^{\rho} + 4\mathcal{H}X^0 \partial_{\mu} X^{\sigma}) 
\end{eqnarray}
For $X^{\mu}$ in Eq.~(\ref{31}), we obtain $\mathcal{X}_{\mu \nu}$ to be
%\iffalse
\begin{subequations}
\begin{eqnarray}
  \mathcal{X}_{00}^N & = & - 2 \big( \frac{2}{a} (\partial_0 (a (\partial_0
  {e}^{(1)} - b^{(1)}))) + (\partial_0 {e}^{(1)} - b^{(1)}) \partial_0 +
  \delta^{ik} (c_k^{(1)} + \partial_k {e}^{(1)}) \partial_i \big)  \big( 2
  \phi^{(1)} \nonumber\\
  &  &- \frac{1}{a} \partial_0 (a (\partial_0 {e}^{(1)} - b^{(1)}))
  \big) + 2 \delta^{ik} (\partial_0  (c_k^{(1)} + \partial_k {e}^{(1)}) )
  (\partial_i b^{(1)} + 2 \nu_i^{(1)} - \partial_0 c_i^{(1)}), \\
  \mathcal{X}_{0 j}^N & = &  \Big( \delta^k_j  \big( \frac{2 \dot{a}}{a}
  (\partial_0 {e}^{(1)} - b^{(1)}) + (\partial_0 {e}^{(1)} - b^{(1)}) \partial_0 +
  \delta^{ik} (c_k^{(1)} + \partial_k {e}^{(1)}) \partial_i + (\partial_0
  (\partial_0 {e}^{(1)} - b^{(1)})) \big)  \nonumber\\
  &  &+ \delta^{ks} (\partial_j (c_s^{(1)} +
  \partial_s {e}^{(1)})) \Big)  (\partial_k b^{(1)} + 2 \nu_k^{(1)} - \partial_0 c_k^{(1)}) - 2
 ( \partial_j  (\partial_0 {e}^{(1)} - b^{(1)}) ) \big( 2 \phi^{(1)}  \nonumber\\
  &  &-  \frac{1}{a} \partial_0 (a (\partial_0 {e}^{(1)} - b^{(1)})) \big)
  + \delta^{kl} (\partial_0  (c_l^{(1)} + \partial_l {e}^{(1)}))  \Big( - 2
  \delta_{jk}  \big( 2 \psi^{(1)} + \frac{\dot{a}}{a} (\partial_0 {e}^{(1)} -
  b^{(1)})   \big) \nonumber\\
  &  &+ 2 \partial_j \partial_k {e}^{(1)} + \partial_j c_k^{(1)} +
  \partial_k c_j^{(1)} + 2h_{j k}^{(1)} \Big), \\
  \mathcal{X}_{kl}^N & = &  \Big(\delta^s_k \delta^t_l  \big( \frac{2 \dot{a}}{a}
  (\partial_0 {e}^{(1)} - b^{(1)}) + (\partial_0 {e}^{(1)} - b^{(1)}) \partial_0 +
  \delta^{ik} (c_k^{(1)} + \partial_k {e}^{(1)}) \partial_i \big) \nonumber\\
  &  & + ((\delta^s_k \delta^{tw} \partial_l + \delta^t_l \delta^{sw}
  \partial_k)  (c_w^{(1)} + \partial_w {e}^{(1)}))\Big)  \Big( - 2 \delta_{st} 
  \big( 2 \psi^{(1)} + \frac{\dot{a}}{a} (\partial_0 {e}^{(1)} - b^{(1)})
  \big) + 2 \partial_s \partial_t {e}^{(1)} \nonumber\\
  &  &+ \partial_s c_t^{(1)} +
  \partial_t c_s^{(1)}  + 2h_{s t}^{(1)} \Big)  + ((\delta^s_k \partial_l + \delta^s_l \partial_k)  (\partial_0 {e}^{(1)}
  - b^{(1)}))  (\partial_s b^{(1)} + 2 \nu_s^{(1)} - \partial_0 c_s^{(1)}) . 
\end{eqnarray}
\end{subequations}

For $X^{\mu}$ in Eq.~(\ref{39}), we have $\mathcal{X}_{\mu \nu}$ in the form of
\begin{subequations}
\begin{eqnarray}
  \mathcal{X}_{00}^S & = & - 2 \Big( 2 \phi^{(1)} + \frac{1}{a}  \int
  \mathrm{d} \eta \{a \phi^{(1)} \} \partial_0 + \delta^{ij}  \int \mathrm{d}
  \eta \big\{ \nu_i^{(1)} + \partial_i b^{(1)} + \frac{1}{a}  \int \mathrm{d}
  \eta' \{a \partial_i \phi^{(1)} \} \big\} \partial_j \Big) \phi^{(1)}
  \nonumber\\
  &  & + 2 \delta^{ij}  \big( \nu_j^{(1)} + \partial_j b^{(1)} + \frac{1}{a}
  \int \mathrm{d} \eta \{a \partial_j \phi^{(1)} \} \big)  (\partial_i
  b^{(1)} + \nu_i^{(1)}), \\
  \mathcal{X}_{0 j}^S & = &  \Bigg(    \delta^k_j\phi^{(1)} +
  \frac{\dot{a}}{a^2} \delta^k_j \int \mathrm{d} \eta \{a \phi^{(1)} \}+ \frac{1}{a} \delta^k_j
  \int \mathrm{d} \eta \{a \phi^{(1)} \} \partial_0 + \delta^{il}\delta^k_j  \int
  \mathrm{d} \eta \big\{ \nu_i^{(1)} + \partial_i b^{(1)} \nonumber\\
  &  & + \frac{1}{a}\int
  \mathrm{d} \eta' \{a \partial_i \phi^{(1)} \} \big\} \partial_l
     + \delta^{lk} \big(\partial_j  \int \mathrm{d} \eta \big\{
  \nu_l^{(1)} + \partial_l b^{(1)} + \frac{1}{a}  \int \mathrm{d} \eta' \{a
  \partial_l \phi^{(1)} \} \big\}\big) \Bigg)  (\partial_k b^{(1)} +
  \nu_k^{(1)}) \nonumber\\
  &  &+ \frac{2}{a} \phi^{(1)} \int \mathrm{d} \eta \{a \partial_j
  \phi^{(1)} \}  + \delta^{ik}  \big( \nu_i^{(1)} + \partial_i b^{(1)} + \frac{1}{a} 
  \int \mathrm{d} \eta \{a \partial_i \phi^{(1)} \} \big)  \Big( - 4
  \psi^{(1)} \delta_{jk} + 4 \partial_j \partial_k {e}^{(1)}  \nonumber\\
  &  & + 2 \partial_j
  c_k^{(1)} + 2 \partial_k c_j^{(1)} + 2 h_{j k}^{(1)} - \frac{2 \dot{a}}{a^2} \delta_{kj}  \int
  \mathrm{d} \eta \{a \phi^{(1)} \}   \nonumber\\
  &  &  + (\delta_{j}^s \partial_k + \delta_{k}^s \partial_j) 
  \int \mathrm{d} \eta \big\{ \nu_s^{(1)} + \partial_s b^{(1)} + \frac{1}{a}
  \int \mathrm{d} \eta' \{a \partial_s \phi^{(1)} \} \big\} \Big), \\
  \mathcal{X}_{kl}^S & = &  \Bigg( \delta^s_k \delta^t_l  \Big( \frac{2
  \dot{a}}{a^2}  \int \mathrm{d} \eta \{a \phi^{(1)} \}+ \frac{1}{a}  \int
  \mathrm{d} \eta \{a \phi^{(1)} \} \partial_0 + \delta^{ij}  \int \mathrm{d}
  \eta \big\{ \nu_i^{(1)} + \partial_i b^{(1)} \nonumber\\
  &  &+ \frac{1}{a}  \int \mathrm{d}
  \eta' \{a \partial_i \phi^{(1)} \} \big\} \partial_j \Big) 
  + \Big( (\delta^s_k \delta^{ti} \partial_l + \delta^t_l
  \delta^{s i} \partial_k) \int \mathrm{d} \eta \big\{ \nu_i^{(1)} + \partial_i
  b^{(1)} + \frac{1}{a}  \int \mathrm{d} \eta' \{a \partial_i \phi^{(1)} \}
  \big\} \Big) \Bigg) \nonumber\\
  &  &  \Big( - 4 \psi^{(1)} \delta_{st} + 4 \partial_s \partial_t
  {e}^{(1)} + 2 \partial_s c_t^{(1)} + 2 \partial_t c_s^{(1)} + 2 h_{s t}^{(1)}\nonumber\\
  &  &   - (\delta_{s}^n   \partial_t + \delta_{t}^n \partial_s)   \int \mathrm{d} \eta
  \big\{ \nu_n^{(1)} + \partial_n b^{(1)} + \frac{1}{a}  \int \mathrm{d}
  \eta' \{a \partial_n \phi^{(1)} \} \big\}  - \frac{2 \dot{a}}{a^2} \delta_{st}  \int \mathrm{d} \eta \{a
  \phi^{(1)} \} \Big) \nonumber\\
  &  &  + \frac{1}{a}  (\partial_s b^{(1)} + \nu_s^{(1)}) 
  \int \mathrm{d} \eta \big\{a (\delta^s_k \partial_l + \delta^s_l \partial_k)
  \phi^{(1)} \big\} . 
\end{eqnarray}
\end{subequations}
\vspace{5mm}\ruledown
\begin{multicols}{2}
\section{Transverse and traceless operator in momentum space}\label{F}

The transverse and traceless operator in configuration space is shown in the
form of
\begin{eqnarray}
  \Lambda^{lm}_{ij} & = & \mathcal{T}^l_i \mathcal{T}^m_j - \frac{1}{2}
  \mathcal{T}_{ij} \mathcal{T}^{lm} \nonumber\\
  & = & (\delta^l_i - \partial^l \Delta^{- 1} \partial_i)  (\delta^m_j -
  \partial^m \Delta^{- 1} \partial_j)\nonumber\\
  & & - \frac{1}{2}  (\delta_{ij} - \partial_i
  \Delta^{- 1} \partial_j)  (\delta^{lm} - \partial^l \Delta^{- 1} \partial^m)
  .  \label{197}
\end{eqnarray}
In momentum space, we could simply let
$\partial_l \rightarrow ik_l$. 
Eq.~(\ref{197}) can be rewritten as
\begin{eqnarray}
  \Lambda^{lm}_{ij} & = & \left( \delta^l_i - \frac{k^l k_i}{\left|
  \text{\bf{k}} \right|^2} \right)  \left( \delta^m_j - \frac{k^m
  k_j}{\left| \text{\bf{k}} \right|^2} \right) \nonumber\\
& & - \frac{1}{2}  \left(
  \delta_{ij} - \frac{k_i k_j}{\left| \text{\bf{k}} \right|^2} \right) 
  \left( \delta^{lm} - \frac{k^l k^m}{\left| \text{\bf{k}} \right|^2}
  \right) \nonumber\\
  & = & (\delta^l_i - n^l n_i)  (\delta^m_j - n^m n_j)\nonumber\\
  & & - \frac{1}{2} 
  (\delta_{ij} - n_i n_j)  (\delta^{lm} - n^l n^m),  \label{198}
\end{eqnarray}
where we have let $n^l \equiv {k^l}/{\left| \text{\bf{k}} \right|}$
for simplicity. 

The 3-dimensional momentum
space can be spanned by a set of normalized
orthogonal bases \{${\epsilon}_i$, $\bar{\epsilon}_i$, $n_i$\} which satisfy
\begin{subequations}
\begin{eqnarray}
    {\epsilon}_i {\epsilon}^i = \bar{\epsilon}_i  \bar{\epsilon}^i = n_i n^i & = & 1,\\
    {\epsilon}_i  \bar{\epsilon}^i = {\epsilon}_i n^i = \bar{\epsilon}_i n^i & = & 0.
\end{eqnarray}
\end{subequations}
%The subscript and superscript are related via ${\epsilon}^i = \delta^{ij} {\epsilon}_i$.
The indices are raised and lowered via  Kronecker delta $\delta^{i}_{j}$, which can be written as
\begin{equation}
  \delta^j_i = {\epsilon}_i {\epsilon}^j + \bar{\epsilon}_i  \bar{\epsilon}^j + n_i n^j . \label{200}
\end{equation}
Substituting Eq.~(\ref{200}) into Eq.~(\ref{198}), we can obtain 
\end{multicols}
\hspace{0mm}\vspace{5mm}\ruleup
\begin{eqnarray}
  \Lambda^{lm}_{ij} & = & ({\epsilon}_i {\epsilon}^l + \bar{\epsilon}_i  \bar{\epsilon}^l)  ({\epsilon}_j {\epsilon}^m +
  \bar{\epsilon}_j  \bar{\epsilon}^m) - \frac{1}{2}  ({\epsilon}_i {\epsilon}_j + \bar{\epsilon}_i  \bar{\epsilon}_j)  ({\epsilon}^l
  {\epsilon}^m + \bar{\epsilon}^l  \bar{\epsilon}^m) \nonumber\\
  & = & {\epsilon}_i {\epsilon}_j {\epsilon}^l {\epsilon}^m + {\epsilon}_i  \bar{\epsilon}_j {\epsilon}^l  \bar{\epsilon}^m + \bar{\epsilon}_i {\epsilon}_j 
  \bar{\epsilon}^l {\epsilon}^m + \bar{\epsilon}_i  \bar{\epsilon}_j  \bar{\epsilon}^l  \bar{\epsilon}^m - \frac{1}{2} 
  ({\epsilon}_i {\epsilon}_j {\epsilon}^l {\epsilon}^m + {\epsilon}_i {\epsilon}_j  \bar{\epsilon}^l  \bar{\epsilon}^m + \bar{\epsilon}_i  \bar{\epsilon}_j {\epsilon}^l
  {\epsilon}^m + \bar{\epsilon}_i  \bar{\epsilon}_j  \bar{\epsilon}^l  \bar{\epsilon}^m) \nonumber\\
  & = & \frac{1}{2}  ({\epsilon}_i  \bar{\epsilon}_j {\epsilon}^l  \bar{\epsilon}^m + \bar{\epsilon}_i {\epsilon}_j {\epsilon}^l 
  \bar{\epsilon}^m + \bar{\epsilon}_i {\epsilon}_j  \bar{\epsilon}^l {\epsilon}^m + {\epsilon}_i  \bar{\epsilon}_j  \bar{\epsilon}^l {\epsilon}^m) +
  \frac{1}{2}  ({\epsilon}_i {\epsilon}_j {\epsilon}^l {\epsilon}^m - {\epsilon}_i {\epsilon}_j  \bar{\epsilon}^l  \bar{\epsilon}^m - \bar{\epsilon}_i 
  \bar{\epsilon}_j {\epsilon}^l {\epsilon}^m - \bar{\epsilon}_i  \bar{\epsilon}_j  \bar{\epsilon}^l  \bar{\epsilon}^m) \nonumber\\
  & = & \frac{1}{\sqrt{2}}  ({\epsilon}_i  \bar{\epsilon}_j + \bar{\epsilon}_i {\epsilon}_j) 
  \frac{1}{\sqrt{2}}  ({\epsilon}^l  \bar{\epsilon}^m + \bar{\epsilon}^l {\epsilon}^m) + \frac{1}{\sqrt{2}} 
  ({\epsilon}_i {\epsilon}_j - \bar{\epsilon}_i  \bar{\epsilon}_j)  \frac{1}{\sqrt{2}}  ({\epsilon}^l {\epsilon}^m - \bar{\epsilon}^l 
  \bar{\epsilon}^m) \nonumber\\
  & = : & {\epsilon}^{\times}_{ij} {\epsilon}^{\times, lm} + {\epsilon}^+_{ij} {\epsilon}^{+, lm},  \label{201}
\end{eqnarray}
\vspace{5mm}\ruledown
\begin{multicols}{2}\hspace{-6.5mm}
where we define
\begin{subequations}
\begin{eqnarray}
  {\epsilon}^{\times}_{ij} & \equiv & \frac{1}{\sqrt{2}}  ({\epsilon}_i  \bar{\epsilon}_j + \bar{\epsilon}_i
  {\epsilon}_j), \\
  {\epsilon}^+_{ij} & \equiv & \frac{1}{\sqrt{2}}  ({\epsilon}_i {\epsilon}_j - \bar{\epsilon}_i  \bar{\epsilon}_j) . 
\end{eqnarray}
\end{subequations}
Here, ${\epsilon}_{ij}^+$ and ${\epsilon}^{\times}_{ij}$ are called the transverse and traceless
polarization tensors. They are widely used to study the  gravitational waves
{\cite{Maggiore:2018sht}}. We note that the third
equal in Eq.~(\ref{201}) is established when
$\Lambda^{lm}_{ij}$ act on the rank-two symmetric tensors.

\end{multicols}
\vspace{10mm}
\begin{multicols}{2}
\bibliography{citation}
\end{multicols}
\end{document}